\documentclass[twocolumn,10pt]{IEEEtran}

\usepackage[utf8]{inputenc}

\usepackage{microtype}
\usepackage{xcolor}
\usepackage{graphicx}
\usepackage{subfigure}
\usepackage{booktabs} 
\usepackage{cite}

\newcommand{\xz}[1]{\textcolor{black}{#1}}

\usepackage{hyperref}

\usepackage{bm}

\usepackage{amsmath}
\usepackage{amssymb}
\usepackage{mathtools}
\usepackage{algorithm}
\usepackage{algorithmic}

\usepackage{lipsum}
\usepackage[capitalize]{cleveref}

\DeclareMathOperator{\Equaldef}{\overset{def}{=}}

\newcommand{\norm}[1]{\left\lVert#1\right\rVert}

\def \< {\langle}
\def \> {\rangle}

\newtheorem{assumption}{Assumption}

\newtheorem{remark}{Remark}
\newtheorem{theorem}{Theorem}
\newtheorem{definition}{Definition}
\newtheorem{corollary}{Corollary}
\newtheorem{lemma}{Lemma}
\newtheorem{example}{Example}

\title{Fast networked data selection via distributed smoothed quantile estimation}

\author{Xu Zhang and Marcos M. Vasconcelos
	\thanks{X. Zhang was supported by the Postdoctoral Fellowship Program of CPSF under Grant Number GZC20232038. M. M. Vasconcelos was partially supported by the Commonwealth Cyber Initiative.}
	\thanks{X.~Zhang is with the School of Artificial Intelligence, Xidian University, Xi'an, China (e-mail: zhang.xu@xidian.edu.cn).}
	\thanks{M. M. Vasconcelos is  with the Department of Electrical Engineering at the FAMU-FSU College of Engineering, Florida State University, USA (e-mail: \texttt{mm22eo@fsu.edu}).}
	
}

\begin{document}

\maketitle

\begin{abstract}
Collecting the most informative data from a large dataset distributed over a network is a fundamental problem in many fields, including control, signal processing and machine learning. In this paper, we establish a connection between selecting the most informative data and finding the top-$k$ elements of a multiset. The top-$k$ selection in a network can be formulated as a distributed nonsmooth convex optimization problem known as quantile estimation. Unfortunately, the lack of smoothness in the local objective functions leads to extremely slow convergence and poor scalability with respect to the network size. To overcome the deficiency, we propose an accelerated method that employs smoothing techniques. Leveraging the piecewise linearity of the local objective functions in quantile estimation, we characterize the iteration complexity required to achieve top-$k$ selection, a challenging task due to the lack of strong convexity. Several numerical results are provided to validate the effectiveness of the algorithm and the correctness of the theory.
\end{abstract}

\section{Introduction}

\begin{figure}[!t]
	\centering
	\includegraphics[width=0.65\columnwidth]{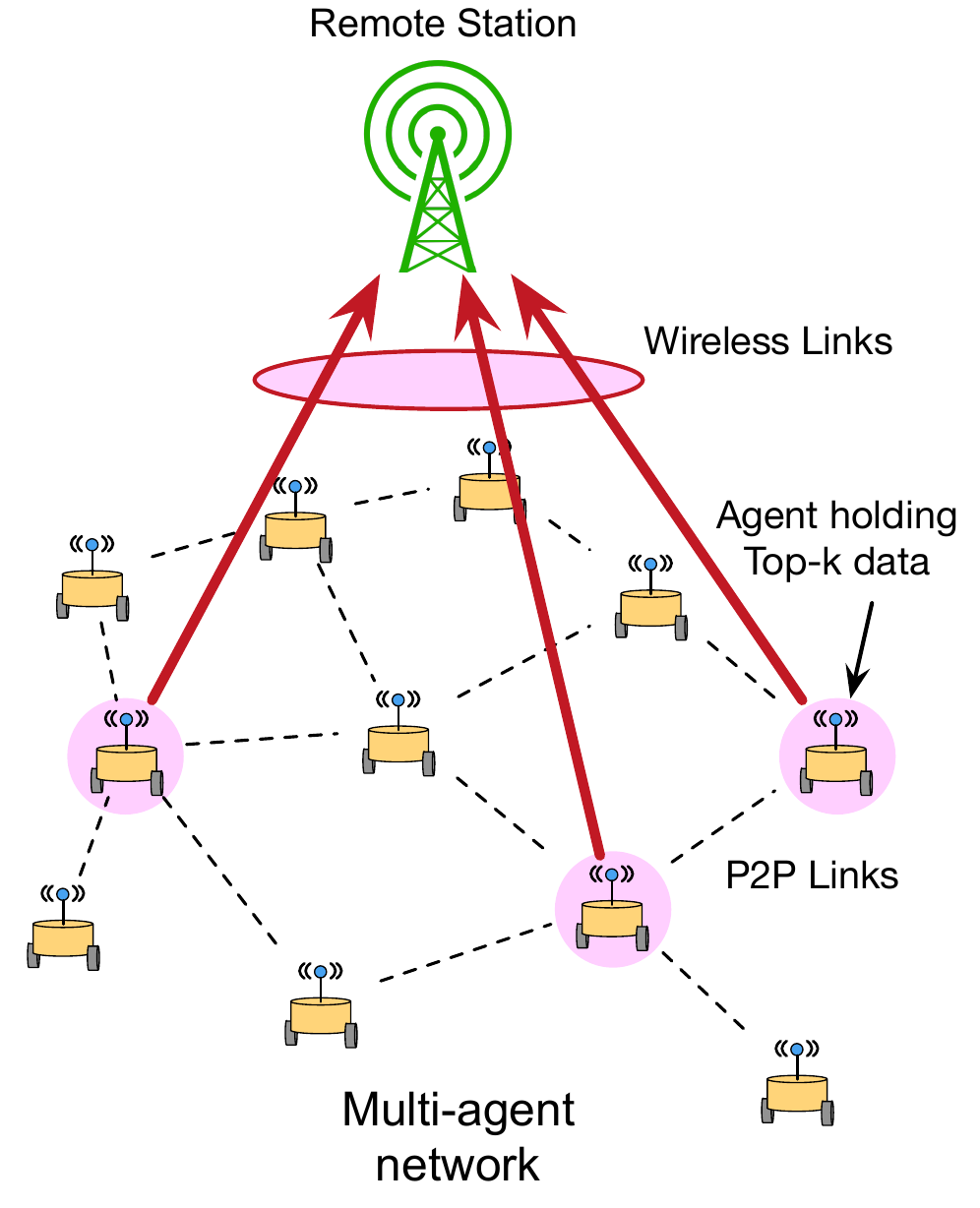}
	\caption{System architecture for the top-$k$ distributed sensor selection problem, where a multi-robot network employs sensors to gather observations, but only the most informative top-$k$ data can be relayed to the remote station via wireless links.} 
	\label{fig:Big Picture}
\end{figure}

Multi-agent networks have been widely used to model many applications such as robotic, sensor and social networks, as well as client-server architectures for distributed machine learning.
With inexpensive sensing devices and storage now readily available, there has been an exponential increase in data generation, leading to the production of vast amounts of data.
However, data processing and wireless communication consume much more power than sensing. Therefore, selecting and transmitting only the most valuable information from a potentially very large collection of random data becomes a fundamental problem in many control, signal processing and machine learning applications \cite{Vasconcelos:2024}. In this paper, we relate this problem to choosing the largest entries in a multiset. While such a selection problem can be easily solved through sorting in centralized systems, a significant challenge arises when dealing with decentralized systems where agents are locally connected over a network through peer-to-peer communication links (see \cref{fig:Big Picture}).






 We address the following problem setting: a dataset is distributed among a potentially large network of $n$ agents interconnected by a local and incomplete communication network, \xz{where agents can only communicate with neighbors through peer-to-peer (P2P) links}. Each agent computes an \textit{informativeness score} for each data point in its local dataset. \xz{For example, in sensor selection based on linear and quadratic estimation models \cite{krause2014submodular, hashemi2019submodular}, informativeness scores are functions of the observation matrices and vectors (c.f. Section II); in the context of polling clients in Federated Learning,  informativeness scores can be a criterion for selection, such as choosing clients with the top-$k$ scores based on the required data upload time to the server \cite{nishio2019client}, the gradient norms of the local loss functions\cite{cho2020client}, or the current value of the local loss functions \cite{marnissi2021client}.} The objective of the agents is to collaborate locally to identify which nodes in the network possess one or more of the top-$k$ most informativeness scores. Subsequently, the selected agents transmit the relevant data points over long-range wireless links to a remote central station for further processing and decision-making.





\subsection{Motivation}

\xz{We would like to solve the top-$k$ selection problem as fast and reliably as possible, and develop a scalable algorithm that accommodates a growing number of agents in the network. 
Despite its apparent simplicity, identifying the top-$k$ largest numbers across a network of interconnected agents poses significant challenges. Numerous studies explore distributed settings in which agents iteratively communicate with a central server to collectively determine the top-$k$ highest informativeness scores \cite{hubschle2016communication, durfee2019practical, shiraishi2020content, malhotra2010exact,liu2016one}. However, relying on a server compromises the robustness and scalability of distributed networks. Moreover, these works entail actual data transmission and require significant storage and communication expenses. 
To address these limitations, distributed sub-gradient methods \cite{Lee:2020, Wang:2018, zhang2021distributed} have been proposed to achieve top-$k$ selection by formulating the problem as a quantile estimation problem. These methods are fully distributed and require smaller storage and communication costs. Nevertheless, they encounter challenges such as slow convergence and lack of analysis regarding iteration complexity. These issues stem from the non-smoothness and non-strong convexity of the objective function. The former renders the application of momentum methods for acceleration, while the latter complicates the analysis of iteration complexity for the iterative variable.}


\subsection{Contributions}


To address the above challenges, this paper proposes an efficient and scalable algorithm for distributed top-$k$ selection based on smoothed distributed quantile estimation. The main contributions are listed below.

\begin{enumerate}
	
	\item By applying smoothing techniques to the local objective functions \cite{nesterov2005smooth, fernandes2021smoothing}, we propose an efficient and scalable distributed top-$k$ selection method. This approach is complemented by integrating it with one of the advanced distributed smooth optimization methods, namely the EXTRA algorithm \cite{2014EXTRA, 2020Revisiting}. Each agent is required to store only two units of memory and transmit just one unit to each agent at every iteration. Moreover, the storage and communication costs remain independent of the number of agents $n$, and the number $k$ for each agent at each iteration, thereby presenting a highly desirable feature in practical applications.
	
	\item  We characterize the iteration complexity of the distributed top-$k$ selection problem for our method, whose expression captures the existing trade-offs between the smoothing parameter, the number $k$, the number of agents, the graph connectivity and the resolution. To give the iteration complexity, the connection between the sequence of objective values and the sequence of iterative variables is established by making full use of the piecewise linear property of the loss function and the uniqueness of the solution.
	
	\item Extensive numerical results substantiate the effectiveness of the proposed method compared to traditional sub-gradient and vanilla message passing methods. Before our current work, distributed sub-gradient algorithms for top-$k$ selection were confined to a small number of nodes. In contrast, our algorithm can scale to very large-scale scenarios and significantly reduce the number of iterations.
	

\end{enumerate}


\subsection{Related Work}



The problem of top-$k$ selection has been an active research topic since the pioneering work of Blum et al. in 1973 \cite{blum1973time}. However, it remains a relevant topic, in which researchers continuously develop algorithms and implementations that harness GPUs to efficiently perform top-$k$ selection on immense datasets \cite{zois2019gpu,zhang2023parallel,Li2024poster}. 

We highlight that the terminology \textit{top-$k$} may correspond to multiple classes of problems in the literature. One such class is called \textit{top-$k$ selective gossip} \cite{ustebay2011efficient,ustebay2012efficiently,ustebay2012top}, where each agent has a vector and engages to reach consensus on the $k$ largest entries of the average of all initial vectors. Despite the apparent similarity in name, \textit{top-$k$ selective gossip} cannot perform \textit{top-$k$ selection}, whose task is to identify within the network the $k$ largest numbers in a multiset. The distributed implementations of top-$k$ selection have been the focus of many fundamental works, including top-$k$ queries \cite{ilyas2008survey,shiraishi2020content,malhotra2010exact,jonsson2012secure,liang2016secure}, top-$k$ monitoring \cite{wu2007top,abbasi2008mote,liu2016one}, gossip-based algorithms \cite{kempe2003gossip,kashyap2006efficient,haeupler2018optimal}, and other message passing algorithms\cite{rajagopal2006universal, kuhn2007tight}. The work reported herein contributes to the state-of-the-art in the top-$k$ selection in networked datasets.

The overwhelming majority of the works in top-$k$ selection, considers the case of aggregating data over a spanning tree for the underlying network. Using message passing algorithms, the top-$k$ data can be consolidated at the root and collected or transmitted to a server. This approach has three drawbacks: (1) It requires the construction or availability of a spanning tree, which may not be available, or the generation may be difficult in very large-scale ad-hoc networks; (2) It requires that the actual data are communicated between nodes in the network, and finally aggregated at the root of the spanning tree, which violates privacy and poses security threats in sensitive applications; (3) It requires communication between any two agents in the network to be noiseless, which is an unrealistic assumption as most modern networks are wireless, and therefore are subject to several forms of communication imperfections such as packet drops and additive noise \cite{liu2011distributed,chen2017critical,zhou2018distributed,yang2020admm,li2022robust,khatana2023noise}.

Our work uses a different approach to solve the top-$k$ selection problem based on distributed convex optimization. A classic result from statistics is that selecting the top-$k$ numbers in a dataset is equivalent to estimating the quantile by minimizing the corresponding \textit{pinball loss} \cite{Koenker:2005}, which has recently been used to solve the distributed top-$k$ problem \cite{Lee:2020}. In \cite{Wang:2018}, the authors noticed that quantile estimation could be solved using distributed optimization. The combination of these results was used in the context of distributed estimation in \cite{zhang2021distributed}, where the connection between the top-$k$ and quantile estimation was formalized, and solved by using a standard distributed subgradient method \cite{nedic2009distributed}. 
One of the big advantages of distributed optimization is that it works under minimal assumptions on the communication network and does not require the construction of a spanning tree. Additionally, there are simple ways of modifying distributed convex optimization algorithms by introducing an extra time scale to handle several types of communication imperfections \cite{doan2020convergence,vasconcelos2021improved,zhang2023top}.
Finally, since our goal is to compute the $k$-th quantile, the actual data are never communicated to any other node. The data remain local, and hence privacy is preserved. One of the drawbacks of current optimization-based algorithms is that they are slow due to the non-smooth nature of the objective function, where diminishing step size is required to guarantee convergence to an exact quantile.



\subsection{Paper Organization}

The rest of this paper is organized as follows. Section \ref{sec: App} introduces two typical applications in sensor subset selection and formulates them as top-$k$ selection problems. Subsequently, Section \ref{sec: SetUp} provides the problem setup by exploring the fundamentals of quantile estimation and modeling top-$k$ selection problem into a distributed quantile estimation problem. Section \ref{sec:ObjFunc} relates the function error of the smoothed objective function and the variable error to the solution in the optimal solution interval. Section \ref{sec: smooth approx} presents two typical smoothing techniques. Section \ref{sec: distributed algorithm} proposes an accelerated distributed quantile algorithm via EXTRA, and provides the iteration complexity. Simulations in Section \ref{sec:simulation} demonstrate the effectiveness of the proposed method. Conclusion and future work are provided in Section \ref{sec:conclution}.
 


\section{Prelude -- Sensor Subset Selection}\label{sec: App}


Sensor subset selection is an NP-hard problem traditionally formulated as an integer programming problem \cite{williamson2011design,krause2014submodular,moon2017static}.
Recent work in this area casts the sensor subset selection as a submodular optimization problem, and by using greedy algorithms, it is possible to obtain good solutions within reasonable time \cite{krause2014submodular, hashemi2019submodular}. In what follows, we illustrate that the problem of subset selection in linear models and quadratic models is equivalent to top-$k$ selection under the so-called T-optimality criterion.  

\subsection{Subset Selection for Linear Models}

Consider a linear model $y_i=\bm{a}_i^T \bm{x} + v_i$, where $\{y_i\}_{i=1}^n \in \mathbb{R}$ are the observations, $\{\bm{a_i}\}_{i=1}^n \in \mathbb{R}^m$ are the observation vectors, $\bm{x} \in \mathbb{R}^m$ is the unknown vector and $\{v_i\}$ are i.i.d. Gaussian noise variables satisfying $v_i \sim \mathcal{N}(0,\sigma_i^2)$ and $\sigma_i$ denotes the standard deviation of $v_i$. 
Consider a scenario where communication limitations dictate that at most $k$ out of $n$ pairs $\{(\bm{a}_i,y_i)\}_{i \in \mathcal{S}}$ can be used to estimate the desired vector $\bm{x}$, where $\mathcal{S} \subset \{1,\ldots,n\}$ with cardinality $|\mathcal{S}|\le k$. Assume that $\bm{x}$ has a Gaussian prior distribution, i.e., $\boldsymbol{x}\sim\mathcal{N}(\bm{0},\mathbf{P})$, where $\mathbf{P}$ denotes its covariance matrix. Following \cite{hashemi2019submodular}, the task of selecting the top-$k$ most informative data points such as to maximize the T-optimality criterion is:
\begin{align} \label{prb: linear1}
	\max_{\mathcal{S}}~~ \text{Tr}(\mathbf{M}_\mathcal{S}^{-1}) -  \text{Tr}(\mathbf{P}^{-1})  ~~\text{  s.t. } \ \ |\mathcal{S}| \le k,
\end{align}
where $\mathbf{M}_\mathcal{S}$ denotes the error covariance matrix \cite{kay2013fundamentals}
\begin{equation} \label{eq: error_covariance}
	\mathbf{M}_\mathcal{S} \Equaldef \left(  \mathbf{P}^{-1} + \sum_{i \in \mathcal{S}} \frac{1}{\sigma_i^2} \bm{a}_i \bm{a}_i^T \right)^{-1}.
\end{equation}

Incorporating \cref{eq: error_covariance} into \cref{prb: linear1}, the problem is reformulated as
\begin{align} \label{prb: linear2}
	\max_{\mathcal{S}}~~ \sum_{i \in \mathcal{S}} \frac{\norm{\bm{a}_i}_2^2}{\sigma_i^2}  ~~\text{  s.t. }\ \  |\mathcal{S}| \le k,
\end{align}
which is equivalent to finding the top-$k$ scores in $\{s_i\}_{i=1}^n$, where 
\begin{equation}
s_i \Equaldef \frac{\norm{\bm{a}_i}_2^2}{\sigma_i^2}.
\end{equation}

\subsection{Subset Selection for Quadratic Models} 

Consider a quadratic measurement model 
\begin{equation}
	y_i=\frac{1}{2}\bm{x}^T \mathbf{A}_i \bm{x} + \bm{b}_i^T\bm{x}+v_i,
\end{equation}
where $\{y_i\}_{i=1}^n \in \mathbb{R}$ are the observations, $\{\mathbf{A}_i\}_{i=1}^n \in \mathbb{R}^{m \times m}$ and $\{\bm{b}_i\}_{i=1}^n \in \mathbb{R}^m$ are known observation matrices and vectors, respectively, $\bm{x} \in \mathbb{R}^m$ is the unknown vector to be estimated and $\{v_i\}$ are i.i.d. Gaussian noise variables satisfying $v_i \sim \mathcal{N}(0,\sigma_i^2)$. We would like to select at most $k$ out of $n$ tuples $\{(\mathbf{A}_i,\bm{b}_i, y_i)\}_{i \in \mathcal{S}}$ to estimate the unknown vector $\bm{x}$. Suppose that $\bm{x} \sim  \mathcal{N}(\bm{0},\mathbf{P})$. Following \cite{hashemi2019submodular}, one way to formulate this problem is to solve:
\begin{align} \label{prb: quadratic1}
	\max_{\mathcal{S}}~~ \text{Tr}(\mathbf{B}_\mathcal{S}^{-1}) -  \text{Tr}(\bm{\Lambda})  ~~\text{  s.t. }\ \ |\mathcal{S}| \le k,
\end{align}
where $\mathbf{B}_\mathcal{S}$ denotes Bayesian Cramér-Rao lower bound according to Theorem 2 in \cite{hashemi2019submodular}
\begin{equation} \label{eq: error_covariance_lowerbounf}
	\mathbf{B}_\mathcal{S}=\left(  \bm{\Lambda} + \sum_{i \in \mathcal{S}} \frac{1}{\sigma_i^2} (\mathbf{A}_i \mathbf{P} \mathbf{A}_i^T+\bm{b_i} \bm{b_i}^T) \right)^{-1},
\end{equation}
and $\bm{\Lambda}$ denotes the Fisher information matrix. 
The optimization problem is reformulated as
\begin{align} \label{prb: quadratic2}
	\max_{\mathcal{S}}~~ \sum_{i \in \mathcal{S}} \frac{1}{\sigma_i^2} \left(\text{Tr}(\mathbf{A}_i \mathbf{P} \mathbf{A}_i^T)+ \norm{\bm{b_i}}_2^2\right)  ~~\text{  s.t. } \ \ |\mathcal{S}| \le k,
\end{align}
which is also equivalent to the top-$k$ selection problem with 
\begin{equation}
s_i\Equaldef \text{Tr}(\mathbf{A}_i \mathbf{P} \mathbf{A}_i^T)+ \norm{\bm{b_i}}_2^2.
\end{equation}

Sensor subset selection is one of the possible applications of the results developed herein. The underlying assumption is that the data is scored using a certain rule. After the scores are obtained, we invoke our algorithms to find in a distributed manner the top-$k$ scores.

\section{Problem Setup} \label{sec: SetUp}



Consider a distributed system with $n$ agents, which interact locally via a connected undirected graph $\mathcal{G}=([n],\mathcal{E})$. Consider a dataset with $N\geq n$ points $\mathcal{D}=\{(x_i,y_i)\}_{i=1}^N$ distributed over the agents in the network. Without loss of generality, we assume that $N=n$, and that each agent has a single data point. The analysis put forth here can easily be extended to a local dataset with more than one point at each node\footnote{The extension to $N\ge n$ is discussed in Appendix \ref{extension:Ngn}.}. 
The $i$-th agent ascribes to its data-point $(x_i,y_i)$ an informativeness score $s_i$, according to an application appropriate metric (c.f. \cref{sec: App}). We highlight that the focus of this work is not on scoring the data, but rather on ranking the scores in a distributed manner.

\vspace{5pt}

\begin{definition}[The $k$-th largest score] \label{def: k-th large} Let $\{s_i\}_{i=1}^n$ denote the collection of all the scores. We arrange the $n$ scores $\{s_i\}_{i=1}^n$ in a new sequence $\{\theta_i\}_{i=1}^n$ in descending order such that $\theta_1 \ge \theta_2 \ge \cdots \ge \theta_n$. Define $\theta_k$ as the $k$-th largest score. We allow for the possibility of repeated scores by denoting $m$ as the number of data points with scores equal to $\theta_k$, and $\overline{m}$ as the number of scores equal to $\theta_k$ whose index in $\{\theta_i\}_{i=1}^n$ is less than or equal to $k$, and $\underline{m}$ as the number of scores equal to $\theta_k$ whose index in $\{\theta_i\}_{i=1}^n$ is strictly larger than $k$.
\end{definition}

\vspace{5pt}

\begin{example}
Consider the multiset $\{2, 2, 5, 1, 2\}$, we obtain the ordered list $\{5, 2, 2, 2, 1\}$, the 2nd, 3rd and 4th largest scores are $2$. 
Furthermore, for $k=3$, we have $\theta_3=2$, $m=3$, $\overline{m}=2$ and $\underline{m}=1$.
\end{example}

\vspace{5pt}

\begin{figure}[!t]
	\centering
	\includegraphics[width=0.6\columnwidth]{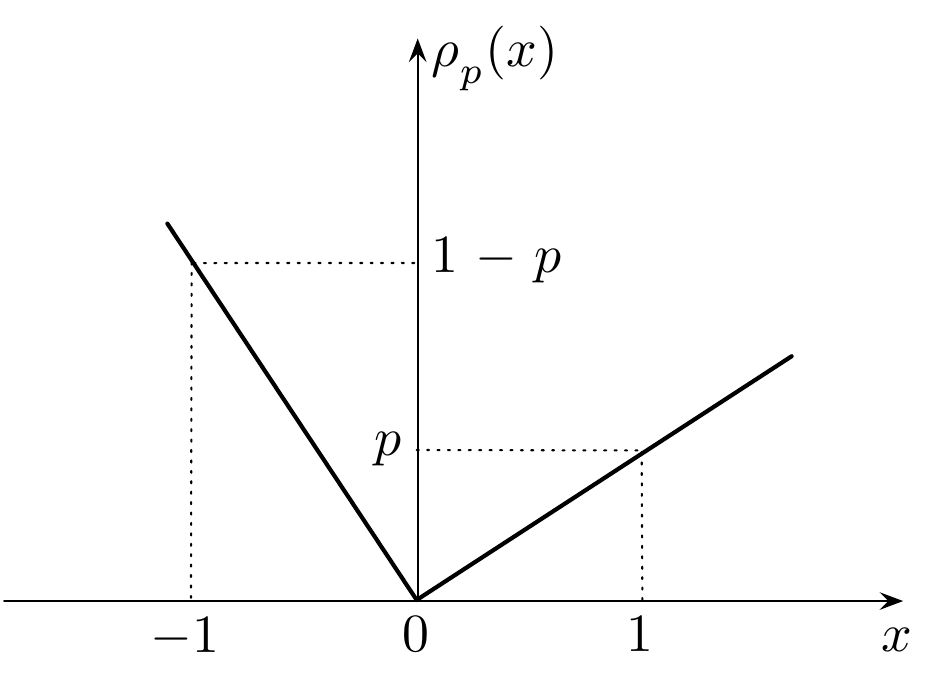}
	\vspace{-0.3cm}
	\caption{Pinball loss function used in quantile estimation. Notice it is neither smooth nor strongly convex.} 
	\label{fig:pinball}
\end{figure}



\xz{Our goal is to determine the top-$k$ largest scores via local communication. We achieve this by obtaining a decentralized algorithm to compute a threshold $T$ between top-$k$ largest score $\theta_k$ and the largest score smaller than $\theta_k$. Once the $i$-th agent has the threshold $T$, it compares $T$ with its score $s_i$. Then the agent knows whether it is holding one of the top-$k$ data points or not. The data point is then transmitted to a remote access point (c.f. \cref{fig:Big Picture}).}


\subsection{Background on Quantile Estimation}
We proceed by relating the computation of the $k$-th largest score $\theta_k$ with sample quantile estimation \cite{Koenker:2005,Lee:2020,zhang2021distributed}. 
Let $F(x)$ be the empirical cumulative distribution function (CDF) of the scores $\{s_i\}_{i=1}^{n}$, i.e,
\begin{equation}
	F(x)= \frac{1}{n} \sum_{i=1}^{n} \mathbf{1}(s_i\le x),
\end{equation}
and $\omega_p$ be the \textit{$p$-th sample quantile} of $\{s_i\}_{i=1}^n$, i.e., 
\begin{equation} \label{def:sample_quantile}
	\omega_p \Equaldef \inf \Big{\{} x\in \mathbb{R} \ \Big| \ F(x) \ge p \Big{\}},
\end{equation}
where $\mathbf{1}(\cdot)$ denotes the indicator function. The following lemma establishes the equivalence between the $k$-largest score $\theta_k$ with the $p$-th sample quantile $\omega_p$ when $p\in (\frac{n-k}{n},\frac{n-k+1}{n})$, which extends the result in Proposition of \cite{zhang2021distributed} from continuous random variables to arbitrary random variables.

\vspace{5pt}


\begin{lemma} \label{lm:k_quantile}
	Let $\{s_i\}_{i=1}^n$ be a sequence of scores
	, then  
	if $p\in (\frac{n-k}{n},\frac{n-k+1}{n})$, we have
	\begin{equation}
		\theta_k=\omega_p.
	\end{equation} 
\end{lemma}

\begin{IEEEproof}
The proof can be found in Appendix \ref{ap:k_quantile}.
\end{IEEEproof}

\vspace{5pt}

From to \cite[Section 1.3]{Koenker:2005}, if $np$ is not an integer, then the $p$-th sample quantile $\omega_p$ is the unique solution of the following quantile estimation problem
\begin{equation} \label{eq:quantile_cvx0}
	\omega_p = \arg \min_{x\in \mathbb{R}} \ \sum_{i=1}^n \rho_p \big(s_i-x \big),
\end{equation}
where 
\begin{equation}
\rho_p(x)\Equaldef \begin{cases}
p\cdot x, & \text{if} \ x\geq0. \\
-(1-p) \cdot 
x, & \text{otherwise},
\end{cases}
\end{equation}
 is the so-called \textit{pinball loss function} (c.f. \cref{fig:pinball}). 

Lemma \ref{lm:k_quantile} implies that if $p\in (\frac{n-k}{n},\frac{n-k+1}{n})$, the $k$-th largest score can be computed as the  solution of the quantile estimation problem:
\begin{equation} \label{eq:quantile_cvx}
	\theta_k = \arg \min_{x\in \mathbb{R}} \  \sum_{i=1}^n \rho_p \big(s_i-x \big) \Equaldef \sum_{i=1}^n f_i(x) \Equaldef f(x),
\end{equation}
where $f_i(x)=\rho_p \big(s_i-x \big)$.

\begin{figure}[!t]
	\centering
	\includegraphics[width=0.85\columnwidth]{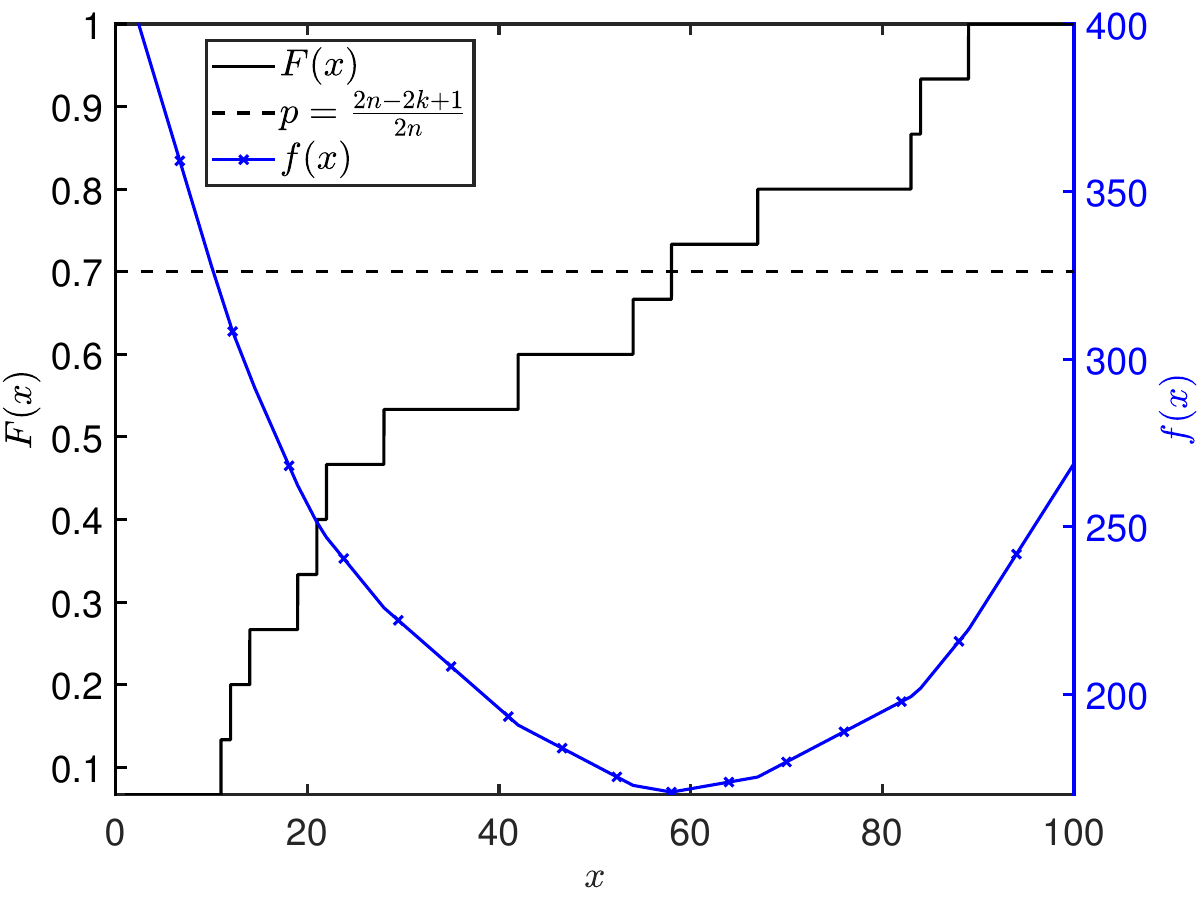}
	\vspace{-0.3cm}
	\caption{Empirical CDF $F(x)$, and its corresponding aggregate pinball loss function $f(x)$ with $n=15$ and $k=5$. The horizontal dotted line denotes the choice of quantile $p$.} 
	\label{fig:ECDF_Objective}
\end{figure}

Notice that $f(x)$ is a \textit{piecewise linear} convex function with a unique minimizer. As shown in Fig. \ref{fig:ECDF_Objective}, we present the empirical CDF $F(x)$, and its corresponding aggregate pinball loss function $f(x)$ with $n=15$ and $k=5$. When $p=\frac{2n-2k+1}{2n}=0.7$, the $p$-th sample quantile corresponds to the minimizer of $f(x)$.

\subsection{Quantile Estimation via Distributed Optimization}

The above problem can be solved using distributed algorithms for non-smooth convex optimization such as the distributed subgradient method \cite{nedic2009distributed},  distributed dual averaging \cite{duchi2011dual} and decentralized alternating direction method of multipliers \cite{makhdoumi2017convergence}, to name a few. However, the aforementioned algorithms tend to be slow due to the non-smoothness of the objective function, which requires either an increased number of communication rounds to achieve convergence or a higher computational cost to address a subproblem in each iteration. Moreover, these algorithms are highly sensitive to noise in the  communication links between the agents \cite{qu2017harnessing,zhang2023top}. 



{ \color{black}
We are interested in characterizing the iteration complexity of top-$k$ selection for a list of $n$ real numbers. Intuitively, the number of iterations required to identify the $k$ largest elements of a list should depend on how close or far apart the elements of the list are, i.e., the search is easier to perform if the elements are distinctively separated.
Therefore, the iteration complexity should increase if the minimum gap between the $k$-th largest score and other scores in the list decrease. To precisely characterize this dependence, we introduce the gap parameter $\Delta(\theta_k)$ to capture the tradeoff between the iteration complexity of our algorithm and the estimation precision, which is defined as follows.

\vspace{5pt}

\begin{definition}[Minimum gap from $\theta_k$] \label{def: values}
	The minimum gap from $\theta_k$, denoted by $\Delta(\theta_k)$, is defined as the minimum absolute difference between $k$-th largest score and other scores in the list:
	\begin{equation}
	\Delta(\theta_k) = \min \Big\{ |s_i-\theta_k| \ \Big| \  s_i \neq \theta_k, \ i = 1,\ldots, n \Big\}.
		\end{equation}
\end{definition}

\vspace{5pt}





Using \cref{def: values}, we proceed by defining the \textit{optimal solution interval} and \textit{optimal threshold interval} as follows.

\vspace{5pt}

\begin{definition}[Optimal solution interval] Let the optimal solution interval, $\mathcal{I}(\theta_k)\subset \mathbb{R}$, be defined as the open interval centered at $\theta_k$ with radius $\Delta(\theta_k)/2$, i.e., 
\begin{equation}
\mathcal{I}(\theta_k)\Equaldef \Big(\theta_k-\frac{\Delta(\theta_k)}{2}, \theta_k+\frac{\Delta(\theta_k)}{2}\Big).
\end{equation}
\end{definition}

\vspace{5pt}

\begin{definition}[Optimal threshold interval] Let the optimal threshold interval, $\mathcal{T}(\theta_k)\subset \mathbb{R}$, be defined as the following open interval: 
\begin{equation} 
\mathcal{T}(\theta_k)\Equaldef \big(\theta_k-\Delta(\theta_k), \theta_k\big).
\end{equation}
\end{definition}

\vspace{5pt}

Notice that for any $s \in \mathcal{I}(\theta_k)$, we can compute a threshold $T$ such that: 
\begin{equation}\label{eq:threshold}
T \Equaldef s-\frac{\Delta(\theta_k)}{2} \in \mathcal{T}(\theta_k).
\end{equation}
Any threshold $T$ within the interval $\mathcal{T}(\theta_k)$ is equally suitable for selecting the top-$k$ elements. This is because all scores below $\theta_k$ are strictly less than $T$, whereas those equal to or exceeding $\theta_k$ are strictly larger than $T$. Therefore, if an iterative algorithm produces a sequence of points $\{w^{t}\}_{t=0}^\infty$ that converges to a value $s$ in the optimal solution interval $\mathcal{I}(\theta_k)$, then we can obtain a desired threshold $T$ using \cref{eq:threshold} that guarantees a correct top-$k$ selection.



}
One key aspect of our analysis is that the definition of \xz{a minimum gap $\Delta(\theta_k)$} allows us to converge to any number within the optimal solution interval, $\mathcal{I}(\theta_k)$ instead of the exact optimal solution $\theta_k$, which is a less stringent convergence condition. In this paper, we design an accelerated algorithm to achieve top-$k$ selection by exploiting the piecewise linearity of the objective functions and the optimal solution interval. We accomplish this task by solving a smoothed distributed quantile estimation problem. Specifically, we will initially establish a correspondence between the (smoothed) optimal value error and the optimization variable error within the optimal solution interval. Subsequently, we will devise appropriate smoothed functions and employ accelerated smooth distributed algorithms to attain our objective. From hereon, we will suppress the dependence of $\theta_k$ from $\Delta(\theta_k)$, to simplify the notation.

\vspace{5pt}

\begin{figure}[!t]
	\centering
	 \includegraphics[width=0.85\columnwidth]{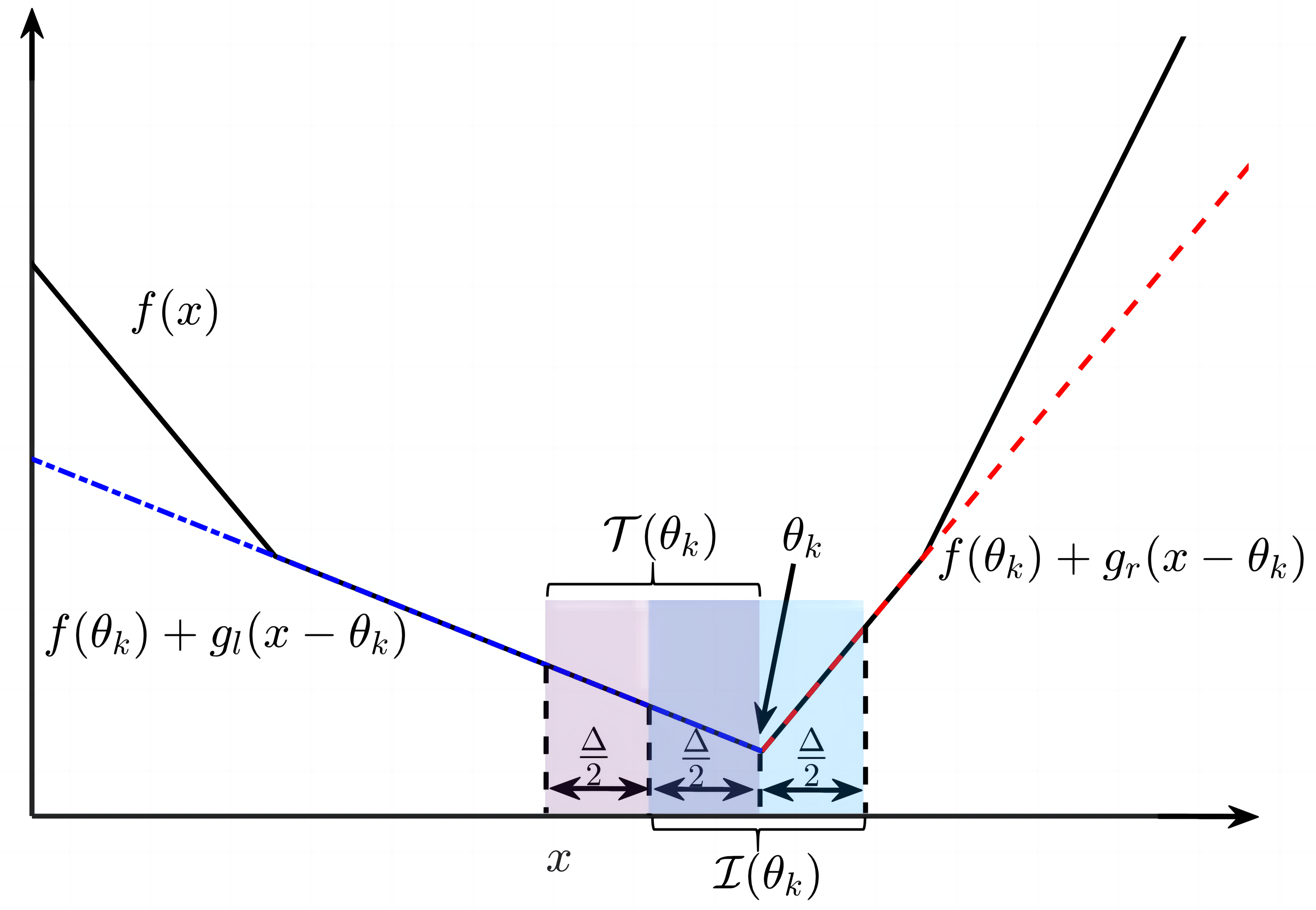}
	\vspace{-0.3cm}
	\caption{Piecewise linear function $f(x)$, its corresponding linear functions $f(\theta_k)+g_l (x-\theta_k)$ and  $f(\theta_k)+g_r (x-\theta_k)$ around the optimal solution $\theta_k$. \xz{The shaded blue area denotes the optimal solution interval $\mathcal{I}(\theta_k)$ and the shaded purple area denotes the optimal threshold interval $\mathcal{T}(\theta_k)$.}} 
	\label{fig:PiecewiseLinearFunction}
\end{figure}


\section{Analysis} \label{sec:ObjFunc}

\xz{In this section, we initially establish a connection between the objective function and variable errors for Problem \eqref{eq:quantile_cvx} . Subsequently, we employ a smooth approximation technique to refine the non-smooth objective function. Next, we elucidate the relationship between the smoothed function and variable errors. Based on the result, we identify the requirements for the smooth approximation and the distributed algorithm, which will be designed in subsequent sections.}

\subsection{Properties of the Original Objective Function}

The first step in our analysis of the distributed quantile estimation problem in \cref{eq:quantile_cvx} is to establish a correspondence between the function and the variable errors. Let $\theta_k$ be the $k$-th largest score among all the nodes in the network, which is found by solving the optimization problem in \cref{eq:quantile_cvx}. Let the \textit{function error} be defined as $f(x)- f(\theta_k)$ and \textit{variable error} be defined as $|x-\theta_k|$. 
In general, for non-smooth functions, we can only obtain the iteration complexity to achieve a predetermined function error, and cannot guarantee the iteration complexity to achieve a predetermined variable error 
\cite{beck2012smoothing}. Fortunately, as shown in Fig. \ref{fig:PiecewiseLinearFunction}, using the piecewise linearity of the objective function and uniqueness of the optimal solution, we can establish a connection between the function error and variable error in the optimal solution interval as follows. 

\vspace{5pt}

 \begin{lemma} \label{lm: f2v}
 	Let $g_r$ and $g_l$ are defined as the right-hand derivative and left-hand derivative of $f$ at $\theta_k$, respectively, i.e.,
\begin{align} 
	g_r = &\lim_{\delta\to 0^+}\frac{f(\theta_k+\delta)-f(\theta_k)}{\delta}, \label{g_r}
	\\
	g_l = &\lim_{\delta\to 0^-}\frac{f(\theta_k+\delta)-f(\theta_k)}{\delta}. \label{g_l}
\end{align}
Suppose $n\cdot p$ is not an integer. If there exists $x$ such that
 	\begin{equation}
 		f(x)- f(\theta_k) \le \min\{g_r, -g_l\} \cdot \frac{\Delta}{2},
 	\end{equation}
 then
  \begin{equation}
 	| x- \theta_k | \le  \frac{\Delta}{2}.
 \end{equation}
 \end{lemma}

 \vspace{5pt}

 \begin{IEEEproof}
The proof can be found in Appendix \ref{ap:f2v}.
\end{IEEEproof}

\vspace{5pt}

\begin{remark}
This lemma also applies to other piecewise linear convex local objective functions and is not exclusive to the top-$k$ problem.
\end{remark}

\vspace{5pt}

 Let $p=\frac{n-k}{n}+\frac{1}{2n}$, then the exact expression for $g_r$ and $g_l$ are given by the following lemma.

\vspace{5pt}

\begin{lemma} \label{ref: derivatives}
	Choosing $p=\frac{n-k}{n}+\frac{1}{2n}$, we have 
	\begin{align}
		g_r= \overline{m}-\frac{1}{2} \text{~~and~~}
        g_l= -\underline{m}-\frac{1}{2}.
    \end{align}
\end{lemma}

 \vspace{5pt}

 \begin{IEEEproof}
The proof can be found in Appendix \ref{ap:derivatives}.
\end{IEEEproof}

\vspace{5pt}

Combining Lemmas \ref{lm: f2v} and \ref{ref: derivatives}, we have the following corollary.

\vspace{5pt}

\begin{corollary} \label{coro: f2v}
	Let $p=\frac{n-k}{n}+\frac{1}{2n}$. If there exists an $x\in\mathbb{R}$ such that
	\begin{equation}
		f(x)- f(\theta_k) \le  \frac{g_m\Delta}{2},
	\end{equation}
then
	\begin{equation}
		| x- \theta_k | \le  \frac{\Delta}{2},
	\end{equation}
where 
\begin{equation}\label{eq:multiplicity_parameter}
g_m\Equaldef \min\big\{\overline{m}-\frac{1}{2}, \underline{m}+\frac{1}{2}\big\}.
\end{equation}
\end{corollary}

\vspace{5pt}

\subsection{Properties of the Smoothed Objective Function}


\xz{Due to the lack of smoothness of the objective function $f(x)$, existing sub-gradient algorithms for quantile estimation suffer from very slow convergence. To improve convergence speed, one effective strategy is to use smoothing, which requires approximating the non-smooth function with a smooth version and subsequently optimizing the resulting function.}

Let function $\tilde{f}_i^h$ be a convex smooth approximation of $f_i$ indexed by a \textit{smoothing parameter} $h$ and  let 
\begin{equation}
\tilde{f}_h(x)\Equaldef \sum_{i=1}^n \tilde{f}_i^h(x), \ \ x\in \mathbb{R}.
\end{equation}
 We say that $\tilde{f}_h$ is a \textit{convex smooth approximation} of $f$. Let $\theta_k^h$ be the minimizer of $\tilde{f}_h(x)$, i.e.,
\begin{equation}
	\theta_k^h \Equaldef \arg \min_{x\in \mathbb{R}} \tilde{f}_h(x).
\end{equation}
In this section, we will establish the relationship between the function error of the smooth approximation function error, denoted as $\tilde{f}_h(x)-\tilde{f}_h(\theta_k^h)$ and variable error $|x-\theta_k|$. Notice that we are interested in minimizing the smooth approximation, but obtaining a guarantee on the variable error of the original non-smooth function.

Before stating our result, we must introduce the following Assumption \ref{assump: smooth approximation} on the smooth approximation. The two typical approximations obtained using Nesterov's  and convolution smoothing techniques satisfy Assumption \ref{assump: smooth approximation} and will be presented in Section \ref{sec: Nesterov’s smoothing} and \ref{sec: Convolution}, respectively.

\vspace{5pt}

\begin{assumption} \label{assump: smooth approximation}
	Let the function $\tilde{f}_h$ denote a convex smooth approximation of $f$ with smoothing parameter $h$. The function $\tilde{f}_h$ is $L_h$-Lipschitz continuous, i.e.,
\begin{equation}
|\tilde{f}_h(x)-\tilde{f}_h(y)| \le L_h|x-y|, \ \ x,y \in \mathbb{R},
\end{equation}
$M_h$-smooth, i.e.,
\begin{align}
|\nabla\tilde{f}_h(x)-\nabla\tilde{f}_h(y)| \le M_h|x-y|, \ \ x,y \in \mathbb{R}
\end{align}
and the approximation error is uniformly bounded, i.e.,
	\begin{align}
		|f(x)-\tilde{f}_h(x)| \le U_h, \ \ x \in \mathbb{R}.
	\end{align} 
Furthermore, the function $\tilde{f}_i^h$ is $\frac{M_h}{n}$-smooth, i.e.,
\begin{equation}
	|\nabla\tilde{f}_i^h(x)-\nabla\tilde{f}_i^h(y)| \le \frac{M_h}{n}|x-y|, \ \ x,y \in \mathbb{R}.
\end{equation}
\end{assumption}

\vspace{5pt}

Theorem \ref{thm: sf2v} provides the theoretical connection between $\tilde{f}_h(x)-\tilde{f}_h(\theta_k^h)$ and $|x-\theta_k|$ within the optimal solution interval, which means that if we can find $x$ satisfying $\tilde{f}_h(x)-\tilde{f}_h(\theta_k^h) < g_m \Delta/4$, then we can obtain the optimal solution interval and the $k$-th largest score.  

\vspace{5pt}

\begin{theorem}[Connection between smoothed function and variable errors] \label{thm: sf2v}
Suppose Assumption \ref{assump: smooth approximation} holds, and let $p=\frac{n-k}{n}+\frac{1}{2n}$.  If there exists $x$ such that the smooth approximation $\tilde{f}_h$ satisfies
\begin{equation}
	\tilde{f}_h(x)-\tilde{f}_h(\theta_k^h) < \frac{g_m \Delta}{4},
\end{equation} 
and a smoothing parameter $h$ such that 
\begin{equation}
	U_h\le \frac{g_m \Delta}{8},
\end{equation}
then we have
	\begin{equation}
	| x- \theta_k | \le  \frac{\Delta}{2}.
	\end{equation}
\end{theorem}

\vspace{5pt}

 \begin{IEEEproof}
Using the triangle inequality, we have
\begin{multline}
	|f(x)-f(\theta_k)| \le |f(x)-\tilde{f}_h(x)|  +|\tilde{f}_h(x)-\tilde{f}_h(\theta_k^h)| 
	\\ +|f(\theta_k)-\tilde{f}_h(\theta_k^h)|.
\end{multline}
From Assumption \ref{assump: smooth approximation}, the first term satisfies
\begin{equation}
	|f(x)-\tilde{f}_h(x)| \le U_h.
\end{equation} 
Combining Assumption \ref{assump: smooth approximation} and the optimality of $\theta_k$ and $\theta_k^h$, we have 
\begin{align}
	f(\theta_k)\le f(\theta_k^h) \le  \tilde{f}_h(\theta_k^h)+U_h,
\end{align}
and
\begin{align}
	f(\theta_k)\ge \tilde{f}_h(\theta_k)-U_h
	\ge \tilde{f}_h(\theta_k^h)-U_h.
\end{align}
Therefore, the third term satisfies
\begin{align}
	|f(\theta_k)-\tilde{f}_h(\theta_k^h)| \le U_h.
\end{align}

Since there exists $h$ such that $U_h\le \frac{g_m \Delta}{8}$ and 
\begin{equation}
	\tilde{f}_h(x)-\tilde{f}_h(\theta_k^h) < \frac{g_m \Delta}{4},
\end{equation}
we get
\begin{equation}
	|f(x)-f(\theta_k)| \le \frac{g_m \Delta}{8}+\frac{g_m \Delta}{4}+\frac{g_m \Delta}{8}=\frac{g_m \Delta}{2}.
\end{equation}

Invoking \cref{coro: f2v}, we obtain
then
	\begin{equation}
		| x- \theta_k | \le  \frac{\Delta}{2}.
	\end{equation}
\end{IEEEproof}

\vspace{5pt}

Based on Theorem \ref{thm: sf2v}, we can  design an accelerated distributed top-$k$ algorithm by accomplishing two tasks:
\begin{itemize}
	\item[(1)] Find a \textit{smooth approximation} $\tilde{f}_h$ that satisfies Assumption \ref{assump: smooth approximation} with 
	\begin{equation}
		U_h\le \frac{g_m \Delta}{8};
	\end{equation}
	\item[(2)] Find a \textit{distributed algorithm} such that 
\begin{equation}
	\tilde{f}_h(w_i^t)-\tilde{f}_h(\theta_k^h) \le \frac{g_m \Delta}{4}, \ \ i=1,\ldots,n
	\end{equation} 
	with $t$ as small as possible, where $w_i^t$ is the local estimate of $\theta_k^h$ computed by agent $i$ at the $t$-th iteration. 
\end{itemize}

In the subsequent sections, we will design smoothing approximations in Section \ref{sec: smooth approx} and provide fast distributed algorithms with corresponding iteration complexity guarantees in Section \ref{sec: distributed algorithm}.


\section{Smooth approximation} \label{sec: smooth approx}

In this section, we consider two smoothing techniques to approximate $f$: The Nesterov’s and convolution smoothing approaches \cite{nesterov2005smooth,beck2012smoothing,fernandes2021smoothing, he2021smoothed}. The rationale here is to introduce the approximations for the pinball loss function in terms of a certain smoothing parameter, and obtain sufficient conditions that guarantee the validity of \cref{assump: smooth approximation}.

\subsection{Nesterov’s smoothing} \label{sec: Nesterov’s smoothing}

Nesterov's smoothing  \cite{nesterov2005smooth,beck2012smoothing} is a systematic way to approximate non-smooth objective functions $f$ by a function with Lipschitz-continuous gradient. Let $h$ be a positive smoothness parameter. The following function is Nesterov's smooth approximation of $f$:
{\color{black}
\begin{equation}
	\tilde{f}_h^{\mathrm{nest}}(x)\Equaldef\sum_{i=1}^n \rho_p^h \big(s_i-x \big),
\end{equation}
where $\rho_p^h(x)$ is the smooth approximation of $\rho_p(x)$
\begin{equation}
	\rho_p^h(x) \Equaldef \max_{ z \in \mathbb{R}} \Big\{ zx - \phi(z) - \frac{h}{2}z^2 \Big\},
\end{equation}
and $\phi(z)$ is the conjugate function of $\rho_p(x)$ 
\begin{equation}
	\phi(z) \Equaldef \max_{x\in \mathbb{R}}\{zx-\rho_p(x)\}.
\end{equation}

Nesterov’s smooth approximation of the score function, $\rho_p^h(x)$, admits a closed-form expression, which is stated as the following result.}

\vspace{5pt}

\begin{lemma} \label{lm:NestSmooth}
The smooth approximation of $\rho_p(x)$ under Nesterov's smoothing, denoted as $\rho_p^h(x)$, is given by     
	\begin{equation}
		\rho_p^h(x) \Equaldef \begin{cases}
			p x -\frac{h}{2} p^2 & \text{if} \ \ x> h p \\
			\frac{x^2}{2 h} & \text{if} \ \  h (p-1) < x \le h p \\
			(p-1) x-\frac{h}{2}(p-1)^2 &\text{if} \ \ x \le h (p-1). \\
		\end{cases}
	\end{equation}
\end{lemma}

\begin{IEEEproof}
The proof can be found in Appendix \ref{append: Pf_Lemma_NS}.
\end{IEEEproof}

\vspace{5pt}
 
Using Nesterov’s smoothing, the optimization problem in \cref{eq:quantile_cvx} can be approximated by 
\begin{equation}
	\theta^{\mathrm{nest}}_h \Equaldef \arg \min_{x\in \mathbb{R}} \tilde{f}_h^{\mathrm{nest}}(x)=\sum_{i=1}^n \rho_p^h \big(s_i-x \big).
\end{equation}

The next Lemma establishes a sufficient condition on the smoothness parameter $h$ such that the Nesterov's approximation satisfies all the conditions in \cref{assump: smooth approximation}.


\vspace{5pt}

\begin{lemma} \label{lm: Nesterov properties}
If the smoothing parameter $h$ satisfies
\begin{equation}
	h \le \frac{g_m \Delta}{4n \max\{p^2,(1-p)^2\}},
\end{equation}	
then the Nesterov's smooth approximation $\tilde{f}_h^{\mathrm{nest}}(\cdot)$ satisfies Assumption \ref{assump: smooth approximation} with
\begin{IEEEeqnarray}{cCl}
	L_h &=& n  \max \{p,1-p \}, \\
	M_h &=& \frac{n}{h}, \\
	U_h &=& \frac{n h }{2} \max\{p^2,(1-p)^2\} \le  \frac{g_m \Delta}{8}.
\end{IEEEeqnarray}
\end{lemma}

\vspace{5pt}

\begin{IEEEproof}
The proof can be found in Appendix \ref{append: Pf_Nest}.
\end{IEEEproof}

\vspace{5pt}

\begin{figure}[t]
	\centering
	\subfigure[]{
	\includegraphics[width=0.8\columnwidth]{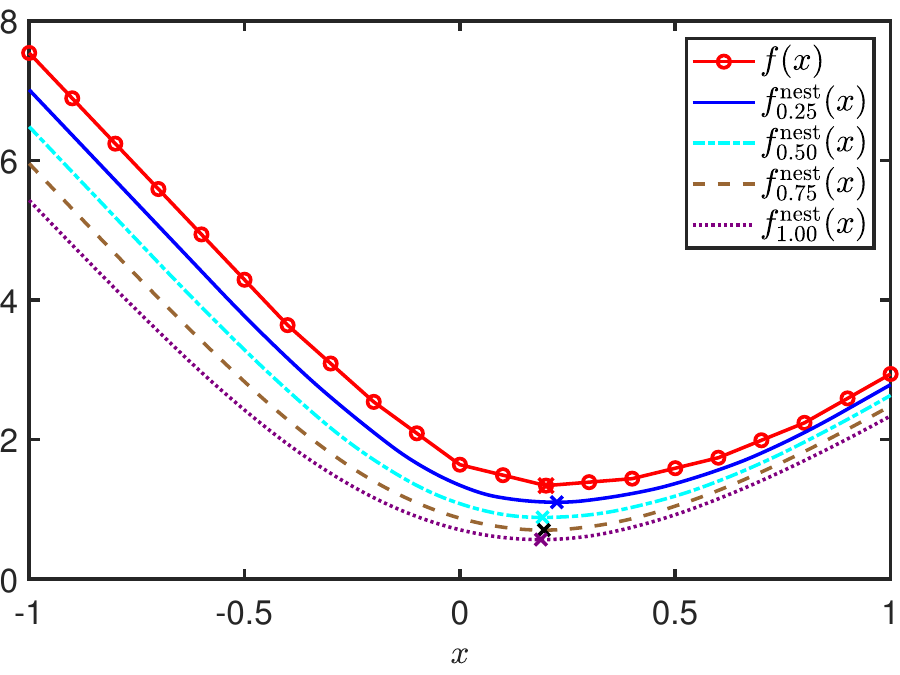}}
	\subfigure[]{
	\includegraphics[width=0.8\columnwidth]{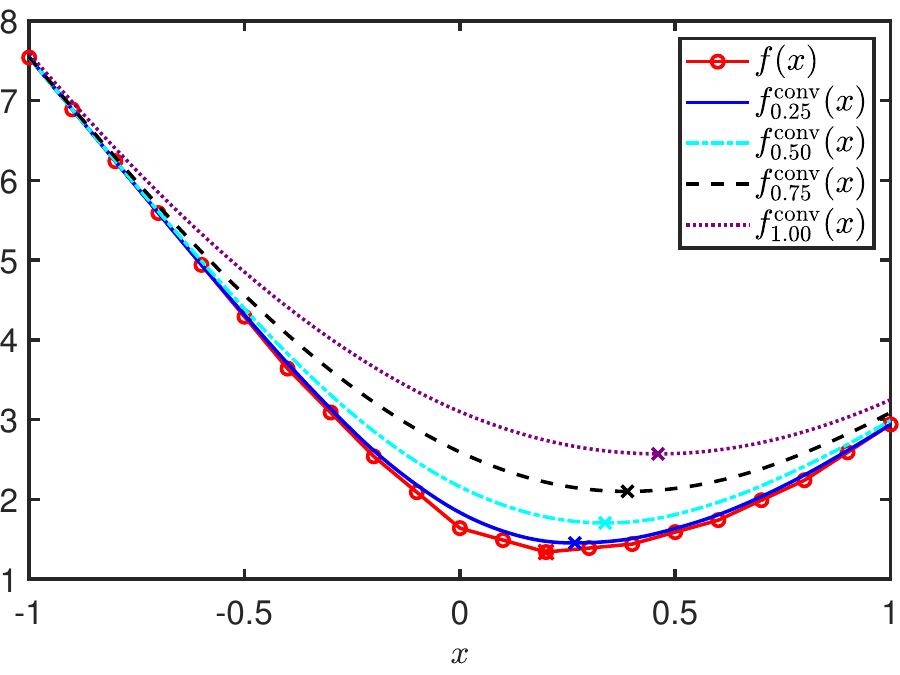}}
	\caption{Examples of smooth approximation for different smoothing parameter $h$ with $n=10$, $k=4$, and $p=0.65$: (a) Nesterov’s smoothing, (b) Convolution smoothing. Here, $f(x)$ denotes the original piecewise linear function, $f_{h}^{\mathrm{nest}}(x)$ denotes the Nesterov's smoothed function, $f_{h}^{\mathrm{conv}}(x)$ denotes the convolution smoothed function, and the marker $\times$ denotes the minimizer of a function.} 
	\label{fig:SmoothApproximation}
\end{figure} 

\subsection {Convolution smoothing} \label{sec: Convolution}

Another way to obtain an approximation with a tunable smoothness parameter use convolutions. By using the so-called \textit{convolution smoothing approach} \cite{fernandes2021smoothing}, the optimization problem in \cref{eq:quantile_cvx} becomes 
\begin{equation}
	\theta_h^{\mathrm{conv}} \Equaldef \arg \min_{x\in \mathbb{R}} \tilde{f}_h^{\mathrm{conv}}(x)=\sum_{i=1}^n l_p^h \big(s_i-x \big),
\end{equation}
where 
\begin{equation}
	l_p^h(x)\Equaldef (\rho_p*K_h)(x)=\int_{-\infty}^{\infty} \rho_p(y) K_h(y-x) {\rm{d}}y,
\end{equation}
where $*$ denotes the convolution operation. The function $K_h(x)$ is a scaling of a \textit{convolution kernel}, $K(x)$, \cite{he2021smoothed}. That is
\begin{equation}
K_h(x)=\frac{1}{h} K \left(\frac{x}{h}\right).
\end{equation}


%

\vspace{5pt}
	
\begin{definition}[Convolution Kernel] \label{assump: Kernel} A convolution kernel function $K(x)$ is a symmetric, nonnegative and bounded function that integrates to one, i.e., 
\begin{align}
& K(x)=K(-x), \ x \in \mathbb{R}, & K(x)\ge 0, \  x \in \mathbb{R}, \\
& \overline{K}=\sup_{x \in \mathbb{R}}K(x) < \infty, & \int_{-\infty}^{\infty} K(x) {\rm d} x =1 .
\end{align} 
\end{definition}


\vspace{5pt}

\begin{example} When $K(x)$ is the uniform kernel with $K(x)=\frac{1}{2} \mathbf{1}(|x|\le 1)$, the smooth approximation of the score function is: 
\begin{equation}
	l_p^h(x) = \begin{cases}
		p x  & \text{if} \ \ x > h \\
		\frac{(x-h)^2}{4 h} +px & \text{if} \ \  -h < x \le h \\
		(p-1) x &\text{if} \ \ x \le -h . \\
	\end{cases}
\end{equation}
\end{example}

\vspace{5pt}

The following lemma provides as sufficient condition on the smoothing parameter $h$ such that $\tilde{f}_h^{\mathrm{conv}}(\cdot)$ satisfies Assumption \ref{assump: smooth approximation} with $U_h\le \frac{g_m \Delta}{8}$.

\vspace{5pt}

\begin{lemma} \label{lm: Convolution properties} For any convolution kernel defined according to \cref{assump: Kernel}, if the smoothing parameter $h$ satisfies
	\begin{equation}
		h \le \frac{g_m \Delta}{8n \max\{p, 1-p\} \int_{-\infty}^{\infty} |t| K(t) {\rm{d}}t},
	\end{equation}	
then the smooth approximation $\tilde{f}_h^{\mathrm{conv}}(\cdot)$ satisfies Assumption \ref{assump: smooth approximation} with
	\begin{IEEEeqnarray}{cCl}
		L_h &=& n  \max \{p,1-p \}, \\
		M_h &=& \frac{n \overline{K}}{h}, \\
		U_h &=&  n h \max\{p,1-p\} \int_{-\infty}^{\infty} |t| K(t) {\rm{d}}t \le  \frac{g_m \Delta}{8}.
	\end{IEEEeqnarray}
\end{lemma}	
\begin{IEEEproof}
	The proof can be found in Appendix \ref{append: pf_ Conv}.
\end{IEEEproof}

\vspace{5pt}

\Cref{fig:SmoothApproximation} shows the curves for $f(x)$ along with its Nesterov's smooth approximation $\tilde{f}_h^{\mathrm{nest}}(x)$ and its convolution smooth approximation $\tilde{f}_h^{\mathrm{conv}}(x)$ for randomly generated data points for different parameters $h$ with $n=10, k=4$ and $p=0.65$. The convolution kernel function is chosen as $K(x)=\frac{1}{2}\mathbf{1}(|x|\le 1)$. First, notice that  $\tilde{f}_{h}^{\mathrm{nest}}(x)$ approximates $f(x)$ from below and $\tilde{f}_{h}^{\mathrm{conv}}(x)$ approximates $f(x)$ from above. Both approximations converge to the original function as $h$ converges to zero.
The main distinction between the two smoothing techniques is the location of the minimizer with respect to the changes in $h$. The minimizers of Nestrov's smooth approximation are close to the solution of quantile estimation, however the distance between the minimizers and the solution does not necessarily decrease as $h$ decreases; On the other hand, the distance between the minimizers of the convolution smooth approximation and the solution of the original function is monotone decreasing as $h$ decreases.

\begin{algorithm}[t]
	\caption{Fast distributed quantile estimation via EXTRA}\label{alg: Extra}
	\begin{algorithmic} 
		\STATE \textbf{Input:} {$ \tilde{f}_{1:n}^{h}, T, w_{1: n}^0, v_{1: n}^0$}
		
		\FOR{$t=0,1,\cdots,T$} 
		\FOR{$i=1,2,\ldots,n$}
		\STATE $w_i^{t+1}=w_i^t-\alpha \Big{(}\nabla  \tilde{f}_i^h(w_i^t)+v_i^t+\frac{\beta}{2} \Big{(} w_i^t- \sum_{j \in \mathcal{N}_i} W_{i, j} w_j^t\Big{)}\Big{)}$  
		\STATE $v_i^{t+1}=v_i^t+\frac{\beta}{2}\Big{(}w_i^{t+1}-\sum_{j \in \mathcal{N}_i} W_{i, j} w_j^{t+1}\Big{)} $
		\ENDFOR
		\ENDFOR
		\STATE \textbf{Output:} { $\bar{w}_{1: n}^{T+1}$ }
	\end{algorithmic}
\end{algorithm}

\section{Complexity of distributed top-$k$ selection} \label{sec: distributed algorithm}

Combining the state-of-the-art algorithm EXTRA \cite{2014EXTRA, 2020Revisiting} and the auxiliary results on the smoothed optimization problem obtained in the previous sections, we are equipped to obtain the iteration complexity of distributed top-$k$ selection. EXTRA is a sophisticated decentralized optimization algorithm for distributed smooth convex problems, which uses the differences of gradients and achieves convergence with a constant step-size.


Let $w_i^t$ denote the local estimate of $\theta_k^h$ computed by node $i$ at the $t$-th iteration and {$\bar{w}_i^T$} be a value generated {from} $\{w_i^t\}_{t=1}^{T}$. Our goal is to obtain an iteration complexity result, i.e., the minimum value of $T$ such that
\begin{equation}
\tilde{f}_h(\bar{w}_i^T)-\tilde{f}_h(\theta_k^h) < \frac{g_m \Delta}{4}, \ \ i=1,\ldots,n.
\end{equation} 
We adopt the EXTRA algorithm and the improved analysis from \cite{2020Revisiting}, where the convergence is given via the running local average
\begin{equation}
	\bar{w}_i^T \Equaldef \frac{1}{T} \sum_{t=1}^{T} w_i^t.
\end{equation}
Let $\bar{\bm{w}} \Equaldef [\bar{w}_1,\ldots,\bar{w}_n]^T$. The algorithm is presented in Algorithm \ref{alg: Extra}, which is written in a primal-dual framework. The primal variable is updated by 
\begin{equation}
w_i^{t+1}=w_i^t-\alpha \Big{(}\nabla  \tilde{f}_i^h(w_i^t)+v_i^t+\frac{\beta}{2} \Big{(} w_i^t- \sum_{j \in \mathcal{N}_i} W_{i, j} w_j^t\Big{)}\Big{)},
\end{equation}
and the dual variable is updated by
\begin{equation}
v_i^{t+1}=v_i^t+\frac{\beta}{2}\Big{(}w_i^{t+1}-\sum_{j \in \mathcal{N}_i} W_{i, j} w_j^{t+1}\Big{)},
\end{equation}
where $\alpha, \beta>0$ are constants, $\mathcal{N}_i$ denotes the set of neighbors that can communicate locally with node $i$ and $W_{i,j}$ is the $(i,j)$-th entry of the weight matrix $\bm{W}$. The readers are referred to \cite{xiao2004fast} for properties and design of weight matrices.

Let $\gamma(\bm{x})\Equaldef (1/n)\sum_{i=1}^n {x}_i$. Following \cite[Lemma 3]{2020Revisiting}, using the fact that each $\tilde{f}_i^h(x)$ is $M_h/n$-smooth, we may rewrite the convergence of function error  $\tilde{f}_h \left(\gamma\left(\bar{\bm{w}}^T\right)\right)-\tilde{f}_h\left(\theta_k^h\right)$ and variable error $ (1/n)\sum_{i=1}^n\left|\bar{ {w}}_i^T-\gamma \left(\bar{\bm{w}}^T\right)\right|^2$.

\vspace{5pt}

\begin{lemma} \label{lm:EXTRA convergence} Suppose that Assumption \ref{assump: smooth approximation} holds and each $\tilde{f}_i^h(x)$ is $M_h/n$-smooth.
 Let $\alpha\Equaldef \frac{1}{2(M_h/n+\beta)}$, $\beta \Equaldef \frac{M_h}{n \sqrt{1-\sigma_2(\bm{W})}}$, and $\bar{\bm{w}}^T\Equaldef \frac{1}{T} \sum_{t=1}^T \bm{w}^t$. If $T \geq \frac{1}{\sqrt{1-\sigma_2(\bm{W})}}$, then the following inequalities hold:
	\begin{equation} \label{eq:EXTRA_R1}
		 \tilde{f}_h \left(\gamma\left(\bar{\bm{w}}^T\right)\right)-\tilde{f}_h\left(\theta_k^h\right) 
		\leq \frac{34}{T \sqrt{1-\sigma_2(\bm{W})}}\left( R_1 M_h+ \frac{R_2 n^2}{M_h} \right),
	\end{equation}
	and
	\begin{equation} 
	\frac{1}{n} \sum_{i=1}^n\left|\bar{{w}}_i^T-\gamma \left(\bar{\bm{w}}^T\right)\right|^2  \leq \frac{16}{T^2\left(1-\sigma_2(\bm{W})\right)}\left(R_1+\frac{R_2 n^2}{M_h^2}\right), \label{neq: EXACT_concensus} 
	\end{equation}
where $\sigma_2(\bm{W})$ is the second largest singular value of $\bm{W}$, $R_1$ is a constant that satisfies $|w_i^{0}-\theta_k^h|^2\le R_1$ and $|\theta_k^h|^2\le R_1$ and $R_2=\max\{p^2, (1-p)^2\}$.
\end{lemma}

\vspace{5pt}

\begin{IEEEproof}
\xz{Eq. \eqref{eq:EXTRA_R1} is derived by assigning $L= \frac{M_h}{n}$ and then multiplying both sides of the first inequality by $n$ in Lemma 3 from \cite{2020Revisiting}. Similarly, Eq. \eqref{neq: EXACT_concensus} is obtained by setting $L= \frac{M_h}{n}$ in the second inequality of Lemma 3 from \cite{2020Revisiting}.}

\end{IEEEproof}

\vspace{5pt}

We provide the exact number of iterations required to enter the optimal solution interval. 

\vspace{5pt}

\begin{theorem}[Exact Complexity of Distributed Top-$k$ Selection]  \label{thm: iterationcomplexity}
Suppose that Assumption \ref{assump: smooth approximation} holds, and $h$ is chosen such that $U_h\le \frac{g_m \Delta}{8}$. Then, we need
\begin{multline}
	T \ge \frac{1}{g_m \Delta  \sqrt{1-\sigma_2(\bm{W})}}  \max \Bigg{\{}  272 \left( R_1 M_h+ \frac{R_2 n^2}{M_h} \right),\\ 32 L_h \sqrt{n} \left( \sqrt{R_1}+\sqrt{R_2}\frac{n}{M_h} \right)
	\Bigg{\}}
\end{multline}
to reach the optimal solution interval using Algorithm \ref{alg: Extra}. 
\end{theorem}

\vspace{5pt}

\begin{IEEEproof}
From Lemma \ref{lm:EXTRA convergence} and Assumption \ref{assump: smooth approximation}, we obtain the following inequalities hold: 
	\begin{IEEEeqnarray}{rcl} 
	 	|&\tilde{f}_h&(\bar{w}_i^T)-\tilde{f}_h(\theta_k^h)|  \nonumber \\
		& \le &  |\tilde{f}_h(\bar{w}_i^T)-\tilde{f}_h(\gamma(\bar{\bm{w}}^T))|  + |\tilde{f}_h(\gamma\left(\bar{\bm{w}}^T\right))-\tilde{f}_h(\theta_k^h)| \nonumber \\
		& \le & L_h |\bar{w}_i^T-\gamma(\bar{\bm{w}}^T)| + \frac{34}{T \sqrt{1-\sigma_2(\bm{W})}}\left( R_1 M_h+ \frac{R_2 n^2}{M_h} \right) \nonumber\\
		& \stackrel{(a)}{\le} &  \frac{4L_h \sqrt{n}}{T \sqrt{1-\sigma_2(\bm{W})}}\left( \sqrt{R_1}+\sqrt{R_2}\frac{n}{M_h} \right) \nonumber\\
		& & \qquad \qquad + \frac{34}{T \sqrt{1-\sigma_2(\bm{W})}}\left( R_1 M_h+ \frac{R_2 n^2}{M_h} \right),
	\end{IEEEeqnarray}
	where inequality $(a)$ follows from \cref{neq: EXACT_concensus} and  $x^2+y^2\le (x+y)^2$, when $x,y \ge 0$.
	
	Thus, if
	\begin{multline}
		T \ge \frac{1}{g_m \Delta  \sqrt{1-\sigma_2(\bm{W})}}  \max \Bigg{\{}  272 \left( R_1 M_h+ \frac{R_2 n^2}{M_h} \right), \\ 32 L_h \sqrt{n} \left( \sqrt{R_1}+\sqrt{R_2}\frac{n}{M_h} \right)
		\Bigg{\}},
	\end{multline}
	we have
	\begin{align}
		|\tilde{f}_h(\bar{w}_i^T)-\tilde{f}_h(\theta_k^h)| \le \frac{g_m\Delta}{4}.
	\end{align} 
	Therefore, invoking Theorem \ref{thm: sf2v},  we have
	\begin{equation} \label{Solution_neighborhood}
		|\bar{w}_i^T-\theta_k| \le \frac{\Delta}{2},~~ i=1,\ldots,n.
	\end{equation} 
\end{IEEEproof}



\subsection{Order of Complexity of Distributed Top-k Selection}

Incorporating the definition of $L_h$, $M_h$ and the largest $h$ satisfying $U_h\le \frac{g_m \Delta}{8}$ in Lemmas \ref{lm: Nesterov properties} and \ref{lm: Convolution properties}, Nesterov's smoothing approach leads to the following iteration complexity to reach the optimal solution interval:
\begin{equation} \label{eq: EXTRA_N}
	\mathcal{O}\bigg{(} \frac{1}{\sqrt{1-\sigma_2(\bm{W})}}  \max \bigg{\{}\frac{n^2}{g_m^2 \Delta^2}, \frac{n^{\frac{3}{2}}}{g_m \Delta},\sqrt{n}\bigg{\}}\bigg{)},
\end{equation}
Similarly, the convolution smoothing approach leads to the following iteration complexity to reach the optimal solution interval:
\begin{multline} \label{eq: EXTRA_C}
	\mathcal{O}\bigg{(} \frac{1}{\sqrt{1-\sigma_2(\bm{W})}}  \max \bigg{\{}\frac{n^2 \overline{K} \int_{-\infty}^{\infty} |t| K(t) {\rm{d}}t }{g_m^2 \Delta^2}, \\ \frac{n^{\frac{3}{2}}}{g_m \Delta},\frac{\sqrt{n}}{\overline{K} \int_{-\infty}^{\infty} |t| K(t) {\rm{d}}t}\bigg{\}}\bigg{)}.
\end{multline}

The expressions above indicate that with the increase of resolution, $\Delta$, and the multiplicity parameter, $g_m$ (c.f. \cref{eq:multiplicity_parameter}), the iteration complexity initially decreases and then stabilizes when $g_m \Delta$ reaches a sufficiently large value. However, increasing the resolution $\Delta$ diminishes the estimation precision of the $k$-th largest value. Moreover, as the network connectivity increases, the parameter $\sigma_2(\bm{W})$ diminishes, resulting in a reduction of the iteration complexity.

Increasing the number of agents contributes to an increase in iteration complexity, primarily because the communication network and its associated diameter expand with the growing number of agents. 

Finally, we discuss the choice of convolution Kernel in \cref{eq: EXTRA_C}. If the resolution parameter $\Delta$ is small, the first term in the max function, that is, 
\begin{equation}
	\frac{n^2 \overline{K} \int_{-\infty}^{\infty} |t| K(t) {\rm{d}}t }{g_m^2 \Delta^2},
\end{equation}
will be the dominant term. Therefore, we should choose a convolution Kernel such that $\int_{-\infty}^{\infty} |t| K(t) {\rm{d}}t$ is minimized. According to \cref{assump: Kernel}, we know $K(t)\le \overline{K}$ for all $t \in \mathbb{R}$. Therefore, the best choice is to place most mass around $t=0$ with magnitude $\overline{K}$, that is, 
 \begin{equation}
 	K^\star(x)= \overline{K}\, \mathbf{1}\left( |x|\le \frac{1}{2\overline{K}} \right).
 \end{equation}

\begin{figure*}[!t]
	\centering
	\subfigure[]{
		\includegraphics[width=0.83\columnwidth]{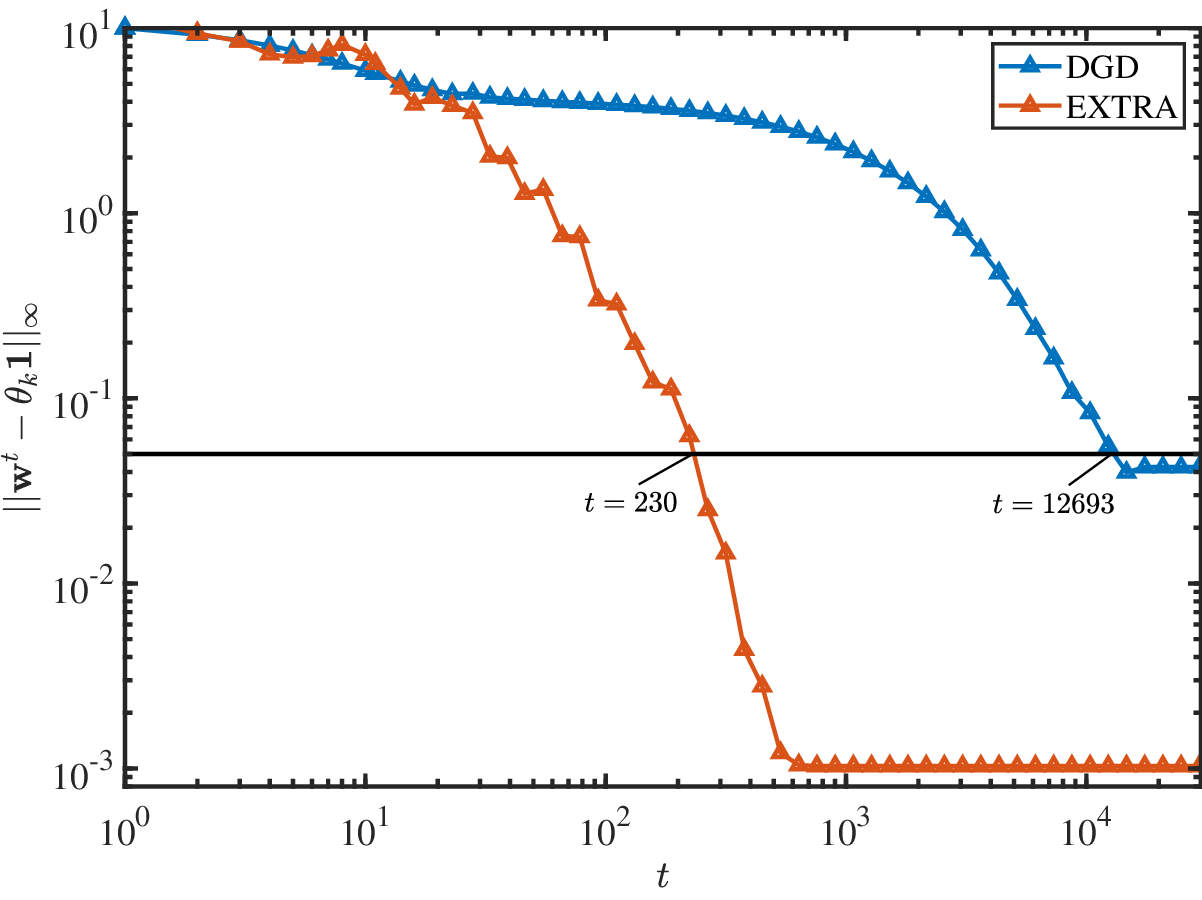}
	}
	\hfil
	\subfigure[]{
		\includegraphics[width=0.8\columnwidth]{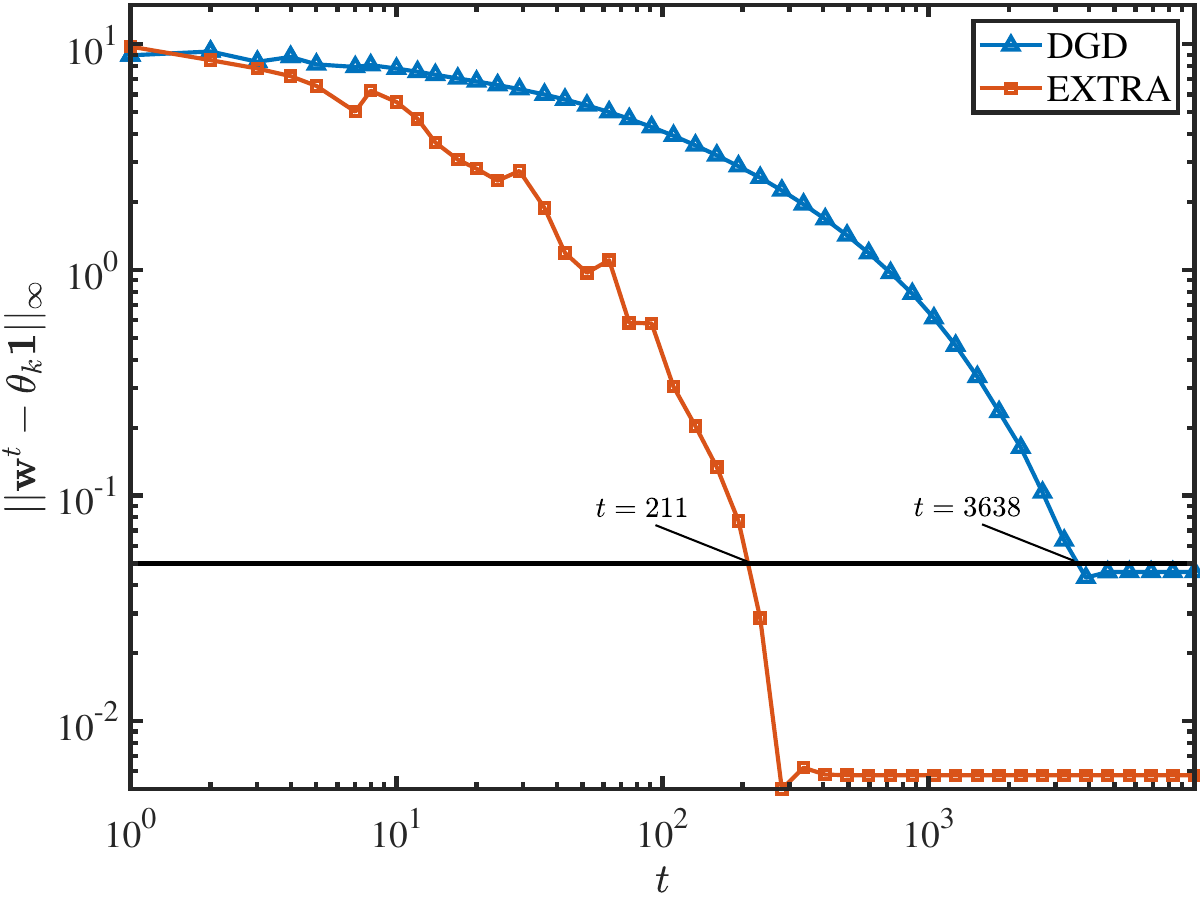}
	}
	\caption{Variable error as a function of the number of iterations with $k=10^3$ and $\Delta=0.1$, (a) $N=n=10^4$; (b) $N=10^5$, $n=10^3$. The black horizontal line denotes the function $||\bm{w}^t-\theta_k \mathbf{1}||_\infty=\frac{\Delta}{2}$.} 
	\label{fig:VarError_vs_iterations_large}
\end{figure*} 
\noindent The implication is that for convolution smoothing the kernel function that minimizes the iteration complexity is the uniform kernel.

\subsection{Fixing the resolution via score quantization}\label{sec:quantization} One of the key parameters in the iteration complexity analysis is the resolution parameter, $\Delta$. We have previously mentioned that this parameter depends on the desired quantile, $\theta_k$, which is unknown. Therefore, the precise iteration complexity varies with each dataset. However, we can adopt a scoring mechanism that guarantees that any two nonidentical consecutive scores are at least $\Delta$ apart. This is accomplished using the following data preconditioning operation.

We process the list of original scores $\{z_i\}_{i=1}^{n}$ by rounding each element using $\Delta>0$ as follows
\begin{equation} \label{eq: roundeddata}
	s_i=\texttt{round}\left(\frac{z_i}{\Delta}\right) \times \Delta, ~~i=1,\ldots,n,
\end{equation}
	where $\texttt{round}(\cdot)$ means rounding to the nearest integer. It is important to notice that by quantizing the scores, information about the true quantile with respect to the original score list is lost. Consequently, it is important to select an appropriate resolution $\Delta$ that strikes a good balance between iteration complexity and quantile estimation accuracy.


\section{Numerical Results} \label{sec:simulation}

In this section, we provide numerical simulations comparing our top-$k$ selection scheme via smoothed quantile estimation via EXTRA with traditional nonsmooth quantile estimation via distributed subgradient descent (DGD) \cite{nedic2009distributed}. We randomly generate scores using the quantization scheme to guarantee a minimum gap $\Delta$ for any $\theta_k$. \xz{Here, we exclusively used convolution smooth approximations. Although the numerical results obtained via Nesterov's smooth approximations are highly comparable, we have omitted them for brevity. Nevertheless, these results can be readily accessed by utilizing the provided code \footnote{The code is available at Github \url{https://github.com/connorzhangxu/DistributedFastTopKSelection}.}. }


The connected graph is an Erdős-Rényi graph, which is generated randomly with $|\mathcal{E}|$ edges. The weight matrix is chosen as $\bm{W}=\bm{I}- \varrho \bm{L}$ according to \cite{xiao2004fast}, where $\varrho=\frac{2}{\lambda_1(\bm{L})+\lambda_{n-1}(\bm{L})}$, $\bm{L}$ is Laplacian matrix of the graph and $\lambda_i(\bm{L})$ denotes the $i$-th largest eigenvalue of a $\bm{L}$. The convolution kernel function is chosen as $K(x)=\frac{1}{2}\mathbf{1}(|x|\le 1)$.   The scores are randomly generated Gaussian variables with variance $\sigma^2$, and then rounded using the scheme in \cref{sec:quantization}. The constraint on the smoothing parameter in Lemma \ref{lm: Convolution properties} directly affects the convergence rate. To accelerate the convergence of the algorithm, we manually adjust the parameter $h$. Following the setting in \cite{2020Revisiting}, we set $\alpha=h$ and $\beta=\frac{1}{h}$. Additionally, we opt for a constant step size manually to accelerate the DGD algorithm towards the optimal solution interval. To guarantee all nodes reach the optimal solution interval, we compare the maximum variable error $||\bm{w}^t-\theta_k \mathbf{1}||_\infty$ with $\Delta/2$ to find the required iterations satisfying $||\bm{w}^t-\theta_k \mathbf{1}||_\infty < \Delta/2$.


\subsection{Convergence rate}


\Cref{fig:VarError_vs_iterations_large} shows the convergence of distributed top-$k$ selection by plotting the variable error as a function of the number of iterations of \cref{alg: Extra}. In \cref{fig:VarError_vs_iterations_large} (a), the agents are collectively estimating the quantile corresponding to the top-$10^3$ scores. The network is sampled from a random graph ensemble with $n=10^4$ nodes and $|\mathcal{E}|=5n$ edges. In this simulation, each agent has a single score, i.e., $N=n=10^4$, which has been randomly drawn from a Gaussian distribution with variance $\sigma^2=10$ and then quantized such that the minimum gap from $\theta_k$ (resolution) is constant, $\Delta=0.1$.
Fig. \ref{fig:VarError_vs_iterations_large} (a) demonstrates that the number of iterations required to identify the top-$k$ for our algorithm is $230$ while for distributed gradient descent (DGD) algorithm is much larger at $12693$. In this case, our algorithm performs approximately $55$ times faster than DGD. This result indicates that we can run the algorithm efficiently in large-scale settings. 

In \cref{fig:VarError_vs_iterations_large} (b), we consider a slightly modified setting. The network is sampled from a random graph ensemble with $n=10^3$ nodes and $|\mathcal{E}|=3n$ edges. \xz{Each agent has a random local dataset with at least one score and the total number of scores across all agents is $N=10^5$}. \Cref{fig:VarError_vs_iterations_large} (b) shows that the number of iterations required to distributedly identify the top-$1000$ scores using our algorithm is $211$, which is approximately $17$ times faster than DGD. Our implementation for top-$k$ selection in this case is very efficient with respect to communication requirements because each agent only communicates a single variable with their neighbors.


\begin{figure}[!t]
	\centering
	\includegraphics[width=0.8\columnwidth]{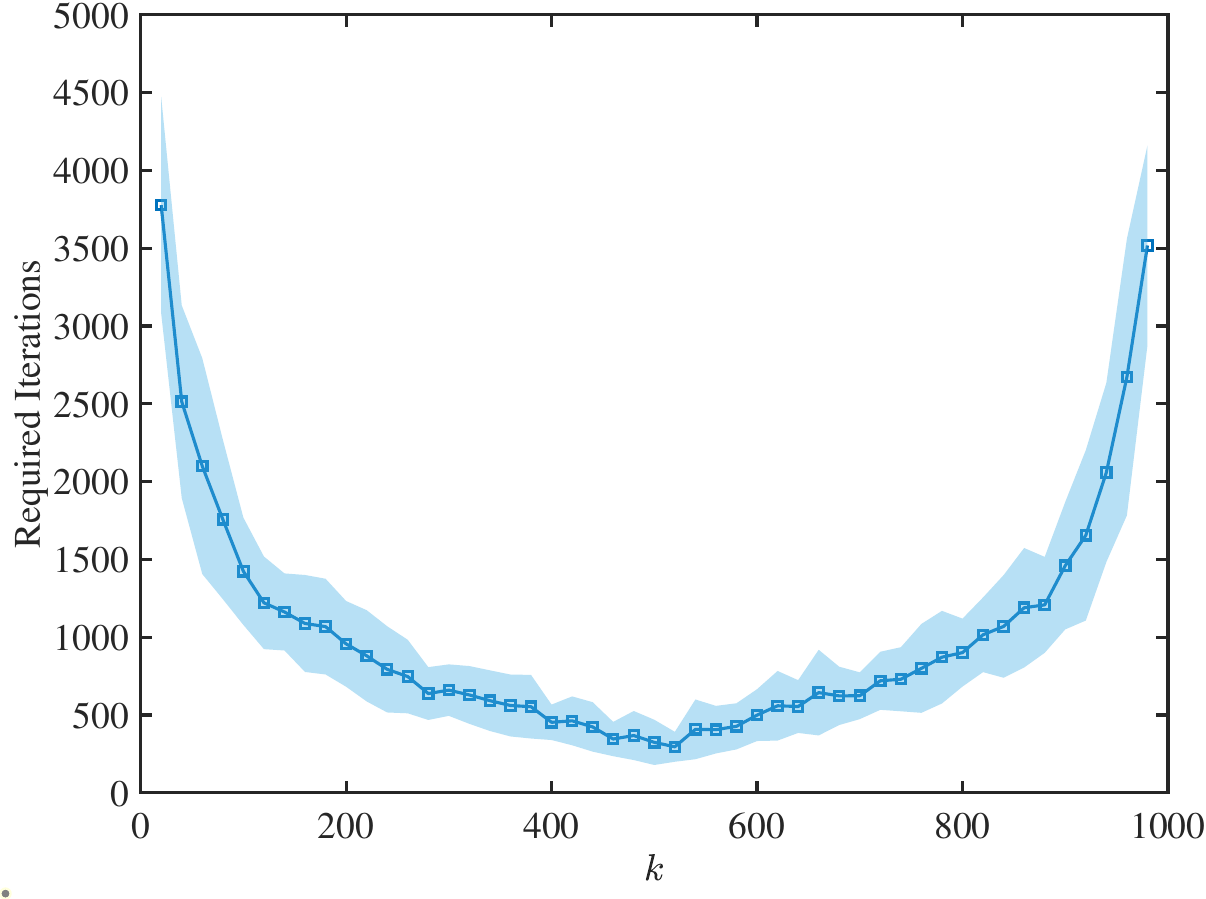}
	\caption{Required iterations to reach the optimal solution interval as a function of $k$ with $n=1000$ and $\Delta=0.1$.  The blue line with a square marker is the average of 100 trials, and the shaded area denotes the standard error.} 
	\label{fig:iterations_vs_k}
\end{figure} 

\subsection{Iteration complexity as a function of desired quantile}

We show the required number of iterations to reach the optimal solution interval as a function of $k$ for $N=n=10^3$ connected by a random graph with $|\mathcal{E}|=3n$ in \cref{fig:iterations_vs_k}. Here we generated a random dataset from a Gaussian distribution with variance $\sigma^2=10$ and quantized them to obtain scores with a resolution of $\Delta=0.1$. We conducted 100 Monte Carlo trials and calculated the average number of required iterations to enter the optimal solution interval. It can be seen that the required iterations are small when $k$ is approximately $n/2$, while they become large when $k$ is close to $1$ and $n$. Therefore, finding the minimum and the maximum elements over the network using \cref{alg: Extra} is harder than finding the median, for example.


\begin{figure}[!t]
	\centering
	\includegraphics[width=0.8\columnwidth]{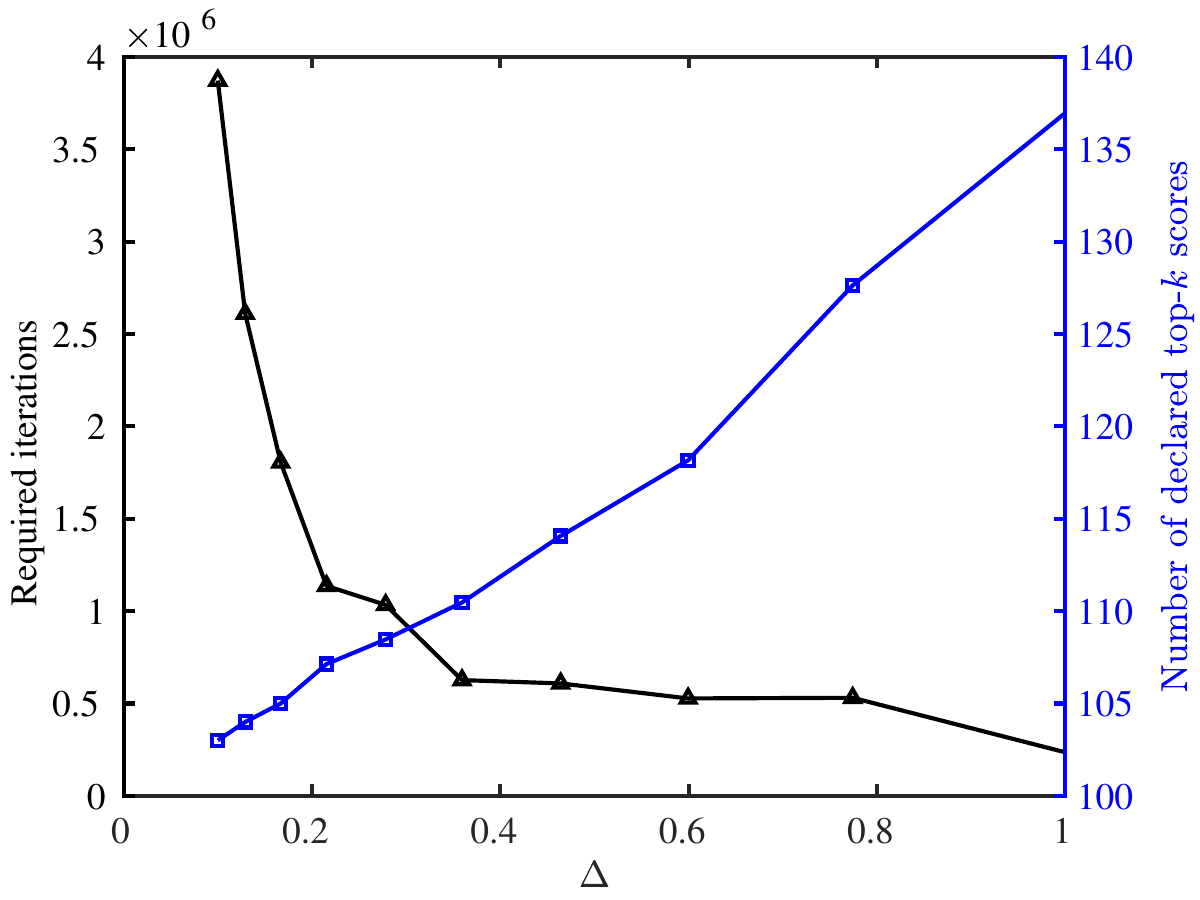}
	\caption{Required iterations to reach the optimal solution interval as a function of $\Delta$ with $n=1000$ and $k=100$. } 
	\label{fig:iterations_vs_Delta_C}
\end{figure} 

\begin{figure}[!t]
	\centering
	\includegraphics[width=0.8\columnwidth]{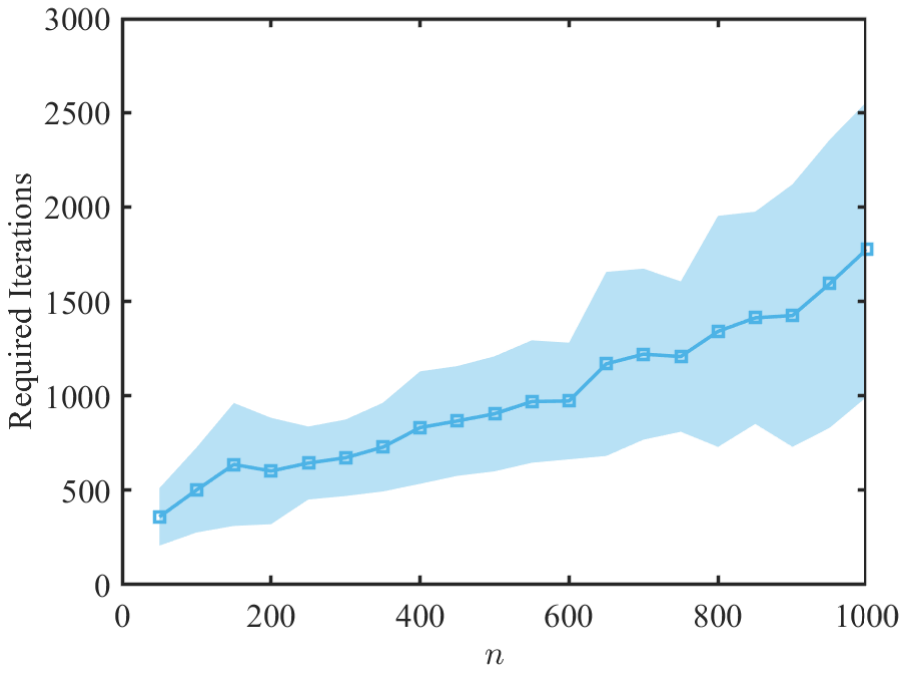}
	\caption{Required iterations to reach the optimal solution interval as a function of $n$ with ${k}/{n}=0.1$ and $\Delta=0.1$. } 
	\label{fig:iterations_vs_n_C}
\end{figure} 

\subsection{Iteration complexity as a function of the quantization gap}

\Cref{fig:iterations_vs_Delta_C} shows the dependence of the total number of  iterations required to reach the optimal solution interval and the number of declared top-$k$ scores as a function of $\Delta$ for $n=10^3$ and $k=10^2$.
We performed 100 Monte Carlo simulations to obtain the results. For each Monte Carlo simulation, we generated random scores $\{s_i\}_{i=1}^n$ and quantized them using different values of $\Delta$. The results show that by increasing $\Delta$, the iterations to reach the optimal solution interval decrease, which coincides with our theoretical results for iteration complexity. In addition, the number of declared top-$k$ scores increases as $\Delta$ increases. This is due to the fact that the number of scores equal to the $k$-th largest value increases as the quantization gap $\Delta$ increases.

\subsection{Iteration complexity as a function of the number of agents}

We also compared the total number of  iterations required to reach the optimal solution interval as a function of $n$ in \cref{fig:iterations_vs_n_C}, where the bold blue line represents the average of 100 trials, and the shaded area denotes the standard deviation. Each trial corresponds to a different set of scores generated randomly sampled from a Gaussian distribution with variance $\sigma^2=100$. To isolate the possible influence of the connectivity graph, we adopted a ring graph with $n$ nodes in this simulation. It is evident that the required iterations increase as the number of nodes increases. However, the average iteration complexity for these randomly generated score lists seems to scale linearly with $n$, which is a desirable feature from the practical perspective. 

\begin{figure}[!t]
	\centering	\includegraphics[width=0.8\columnwidth]{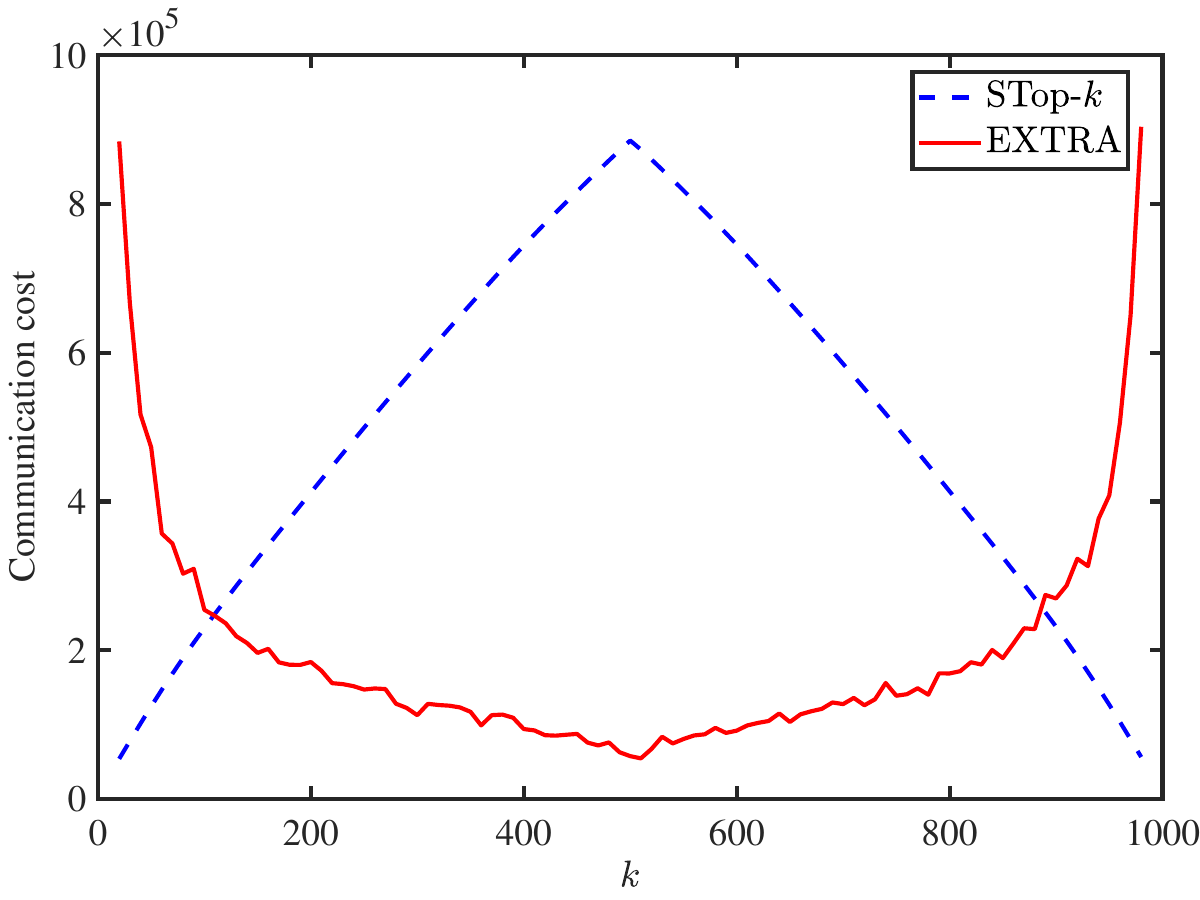}
	\caption{Communication cost to achieve top-$k$ selection as a function of $k$ with $N=1000$, $n=100$ and $\Delta=0.1$. } 
	\label{fig:CommunicationCost}
\end{figure} 

\subsection{Communication cost}

Finally, we compare our distributed top-$k$ selection method against a straightforward and intuitive \xz{message-passing} algorithm outlined in \cite{zhang2023top}. The simple strategy operates as follows:  
Each node maintains a list of at most $k$ scores with their corresponding indices in its memory. At iteration $t$, each node transmits this list to its neighboring nodes. At iteration $t + 1$, each node updates its list by selecting the top-$k$ scores and discarding the rest. Subsequently, each node sorts its list and repeats this process iteratively. To operate more efficiently, when $k \le n/2$, each node selects the top-$k$ scores; when  $k > n/2$, each node chooses the bottom-$k$ scores. We refer to this scheme as \textit{Simple Top-$k$} (STop-$k$) method. \Cref{fig:CommunicationCost} shows the communication cost is simulated for a fixed graph with $|\mathcal{E}|=3n$ edges using the STop-$k$ and our method using EXTRA. Here, we have used $N=10^4$, $n=10^2$ and $\Delta=0.1$. The communication cost denotes the number of transmitted scalars, without taking into account the cost for transmitting the indices in the STop-$k$ method. The results are averaged over 100 trials. From this experiment, it is evident that EXTRA outperforms STop-$k$ for $k$ between 100 and 900, specially when $k$ is approximately $500$. This difference arises due to the fact that the communication cost of STop-$k$ scales with $k$, whereas our method using EXTRA relatively insensitive to $k$ except when used to compute extrema (e.g. minima and maxima). Therefore, using our algorithm is advantageous for a wide range of applications, in which we are interested in determining the top-$k$ scores when $k$ is not small compared to the total number of scores, $N$.

\section{Conclusion and Future Work} \label{sec:conclution}

\xz{Top-$k$ selection algorithms have been studied in many forms for decades and remain relevant in technology, specially in distributed systems. Our distributed top-$k$ algorithm is based on distributed optimization and offers a versatile solution to determine the top-$k$ largest entries in a networked dataset. This framework is applicable in various fields such as wireless sensor networks, signal processing, and machine learning. Unlike existing methods relying on spanning trees, our approach employs distributed optimization and is adaptable to handle noise, packet drops, and other communication imperfections. Leveraging the properties of our local objective functions, we introduced an accelerated algorithm based on smoothing techniques, and expressions for its iteration complexity. As a side product, our method promotes privacy by avoiding data transmission, as nodes estimate a single common threshold to determine whether they are holding or not a top-$k$data-point. Simulations demonstrated its superiority over Distributed Gradient Descent and a simple aggregation-based method called STop-$k$. Future work involves addressing communication imperfections and enabling asynchronous implementation using Gossip mechanisms.}



\bibliographystyle{IEEEtran}

\bibliography{IEEEabrv,refs}

\appendices
\section{Proof of Lemmas}

\subsection{Proof of Lemma \ref{lm:k_quantile}}\label{ap:k_quantile}
	According to Definition \ref{def: k-th large}, the number of scores equal to $\theta_k$ is $m$, the number of scores larger than  $\theta_k$ is $k-\overline{m}$ and the number of scores smaller than  $\theta_k$ is $n-k-\underline{m}$, where $\overline{m}+\underline{m}=m$, $\overline{m}\ge 1$ and $\underline{m}\ge 0$. So we have
	\begin{align}
		\frac{1}{n} \sum_{i=1}^{n} \mathbf{1}(s_i < \theta_k) =\frac{n-k-\underline{m}}{n}<p
	\end{align}
	and
	\begin{align}
		\frac{1}{n} \sum_{i=1}^{n} \mathbf{1}(s_i \le \theta_k) =\frac{n-k+\overline{m}}{n} >p.
	\end{align}
	Together with the definition of \textit{$p$-th sample quantile} in \eqref{def:sample_quantile}, we obtain $\theta_k=\omega_p$.

\subsection{Proof of Lemma \ref{lm: f2v}}\label{ap:f2v}

According to Section 1.3 in \cite{Koenker:2005}, when $np$ is not an integer, the solution of Problem 
\eqref{eq:quantile_cvx} is unique. Together with the fact that $f(x)$ is a piecewise linear function, we can show that if 
\begin{align}
	|f(x)-f(\theta_k)| \le \min\{g_r,-g_l\} \cdot \frac{\Delta}{2},
\end{align} 
then we can guarantee $	| x- \theta_k | \le  \frac{\Delta}{2}$.

\subsection{Proof of Lemma \ref{ref: derivatives}}\label{ap:derivatives}
	By direct superposition the left-hand and right-hand derivatives of $\{\rho_p \big(s_i-x \big)\}_{i=1}^n$ at $\theta_k$, we obtain
	\begin{align}
		g_r = & \sum_{i=1}^{n}(-p \mathbf{1}(s_i > \theta_k)+(1-p) \mathbf{1}(s_i \le \theta_k)) \\
		= & -(k-\overline{m})p+(n-k+\overline{m})(1-p)\\
		=&n-k+\overline{m}-np  
   	\end{align}
		and
		\begin{align}
		g_l = & \sum_{i=1}^{n}( -p \mathbf{1}(s_i \ge \theta_k)+ (1-p) \mathbf{1}(s_i<\theta_k)) \\
		= & [-(k+\underline{m})p+(n-k-\underline{m})(1-p)]\\
		= & n-k-\underline{m}-np.
	\end{align}
	
	Incorporating $p=\frac{n-k}{n}+\frac{1}{2n}$ obtains
	\begin{equation}
		g_r= \overline{m}-\frac{1}{2}, ~~
		g_l= -\underline{m}-\frac{1}{2}.
	\end{equation}
	

\subsection{Proof of Lemma \ref{lm:NestSmooth}} \label{append: Pf_Lemma_NS}

The conjugate function $\phi$ of $\rho_p$ is equivalent to
\begin{equation}
	\phi(z)=\max_{x\in \mathbb{R}}\{zx-x(p-\mathbf{1}(x<0))\}.
\end{equation}
By simple calculation, we obtain
\begin{equation} \label{func: phi}
	\phi(z)=0, \, p-1 \le z\le p.
\end{equation}
Notice that 
\begin{equation}
	\rho_p^h(x)=\max_{ z \in \mathbb{R}} \Big\{ zx - \phi(z) - \frac{h}{2}z^2 \Big\}
\end{equation}
is the conjugate of $\phi(z)+\frac{h}{2}z^2$. Since $\phi(z)+\frac{h}{2}z^2$ is $h$-strongly convex, we have $\rho_p^h(x)$ is $\frac{1}{h}$-smooth.
 Using \eqref{func: phi} gets
\begin{align}
	&\rho_p^h(x) =\max_{ p-1 \le z\le p } \Big\{ zx - \frac{h}{2}z^2 \Big\} \\ =&\begin{cases}
		p x -\frac{h}{2} p^2 & \text{if} \ \ x> h p \\
		\frac{x^2}{2 h} & \text{if} \ \  h (p-1)\le x \le h p \\
		(p-1) x-\frac{h}{2}(p-1)^2 &\text{if} \ \ x < h (p-1). \\
	\end{cases}
\end{align}

\subsection{Proof of Lemma \ref{lm: Nesterov properties}} \label{append: Pf_Nest}
By taking the first-order and second-order derivative of $\rho_p^h(x)$, we get
\begin{align}
	\nabla \rho_p^h(x) =\begin{cases}
		p  & \text{if} \ \ x> h p \\
		\frac{x}{h} & \text{if} \ \  h (p-1)\le x \le h p \\
		(p-1)  &\text{if} \ \ x < h (p-1) \\
	\end{cases}
\end{align}
and
\begin{align}
	\nabla^2 \rho_p^h(x) =\begin{cases}
		\frac{1}{h} & \text{if} \ \  h (p-1)\le x \le h p \\
		0  &\text{otherwise.} 
	\end{cases}
\end{align}
For all $x \in \mathbb{R}$, we obtain
	\begin{align}
		|\nabla \rho_p^h(x)| \le \max \{p,1-p \} \text{~~and~~}
		0 \le \nabla^2 \rho_p^h(x) \le \frac{1}{h}.
	\end{align}
Using the fact that $\tilde{f}_h^{\mathrm{nest}}(x)=\sum_{i=1}^n \rho_p^h \big(s_i-x \big)$, we have
	\begin{align}
	|\nabla \tilde{f}_h^{\mathrm{nest}}(x))| \le n\max \{p,1-p \} \text{~and~}
	0 \le \nabla^2 \tilde{f}_h^{\mathrm{nest}}(x) \le \frac{n}{h}.
\end{align}
So $\tilde{f}_h^{\mathrm{nest}}(\cdot)$ is a convex, $L_h=n  \max \{p,1-p \}$-Lipschitz continuous, $M_h=\frac{n}{h}$-smooth function.
By basic calculations, we have
\begin{align}
	\rho_p^h(x)- \rho_p(x) =\begin{cases}
		\frac{h}{2} p^2 & \text{if} \ \ x> h p \\
		px-\frac{x^2}{2 h} & \text{if} \ \  0< x \le h p \\
		(p-1)x-\frac{x^2}{2 h} & \text{if} \ \  h (p-1)\le x \le 0 \\
		\frac{h}{2}(p-1)^2 &\text{if} \ \ x < h (p-1), \\
	\end{cases}
\end{align}
and 
\begin{equation} \label{smooth_approximation_inequality1}
	0 \le 	\rho_p^h(x)- \rho_p(x) \le 	 \frac{h }{2} \max\{p^2,(1-p)^2\}.
\end{equation}

Using the fact that $\tilde{f}_h^{\mathrm{nest}}(x)=\sum_{i=1}^n \rho_p^h \big(s_i-x \big)$ again yields
	\begin{equation} \label{smooth_approximation_inequality2}
		\tilde{f}_h^{\mathrm{nest}}(x) \le 	f(x) \le 	\tilde{f}_h^{\mathrm{nest}}(x)+ \frac{n h }{2} \max\{p^2,(1-p)^2\},
	\end{equation}
Therefore, $|f(x) - 	\tilde{f}_h^{\mathrm{nest}}(x)| \le U_h \Equaldef \frac{n h }{2} \max\{p^2,(1-p)^2\}$.

\subsection{Proof of Lemma \ref{lm: Convolution properties}} \label{append: pf_ Conv}
	The first-order and second-order derivative of $\tilde{f}_h^{\mathrm{conv}}(x)$ are
	\begin{align}
		\nabla \tilde{f}_h^{\mathrm{conv}}(x) &= \sum_{i=1}^n \left[\mathcal{K}_h(x-s_i)-p \right], \label{eq: 1st derivative}\\
		\nabla^2 \tilde{f}_h^{\mathrm{conv}}(x) &= \sum_{i=1}^n K_h(x-s_i), \label{eq: 2nd derivative}
	\end{align}
	where
	\begin{align}
		\mathcal{K}(x)&=\int_{-\infty}^{x} K(y) {\rm{d}}y \text{~~and~~} \mathcal{K}_h(x)=\mathcal{K}\left( \frac{x}{h} \right).
	\end{align}
	
	According to \cref{assump: Kernel}, we have 
	\begin{align}
		0 \le &\mathcal{K}_h(x) \le 1,~ \forall x \in \mathbb{R}, \\
		0 \le &K_h(x) \le \frac{\bar{K}}{h},~ \forall x \in \mathbb{R}
	\end{align}

	Together with Eqs. \eqref{eq: 1st derivative} and \eqref{eq: 2nd derivative}, we get
	\begin{align}
		|\nabla \tilde{f}_h^{\mathrm{conv}}(x)| &\le n \max\{p,(1-p)\} \Equaldef L_h, \\
		0 \le \nabla^2 \tilde{f}_h^{\mathrm{conv}}(x) &\le \frac{n \overline{K}}{h} \Equaldef M_h.
	\end{align}
	So $\tilde{f}_h^{\mathrm{conv}}(\cdot)$ is convex, $L_h$-Lipschitz continuous, $M_h$-smooth.
	Next, let's give the bound of $|f(x)-\tilde{f}_h^{\mathrm{conv}}(x)|$. First of all, using variable substitution and Jenson's inequality yield
	\begin{align}
		l_p^h(x)&=\int_{-\infty}^{\infty} \rho_p(y) K_h(y-x) {\rm{d}}y \\
		&= \int_{-\infty}^{\infty} \rho_p(x+t) K_h(t) {\rm{d}}t \\
		&\ge \rho_p\left(x+\int_{-\infty}^{\infty} t K_h(t) {\rm{d}}t \right)\\
		&= \rho_p(x), \forall x \in \mathbb{R},
	\end{align}
	where the last equality uses the even function property of $K_h(t)$. So we have
	\begin{align} \label{eq: conv_lower_bound}
		\tilde{f}_h^{\mathrm{conv}}(x)-f(x)= \sum_{i=1}^n \left[ l_p^h \big(s_i-x \big)-\rho_p \big(s_i-x \big)\right] \ge 0.
	\end{align}
	
	Using $\int_{-\infty}^{\infty} K_h(t) {\rm{d}}t=1$ yields
	\begin{align}
		l_p^h(x)-\rho_p(x) = & \int_{-\infty}^{\infty} \big(\rho_p(x+t) - \rho_p(x) \big) K_h(t) {\rm{d}}t,\\
		\le& \int_{-\infty}^{\infty} \big|\rho_p(x+t) - \rho_p(x) \big| K_h(t) {\rm{d}}t \\
		\le& \max\{p,1-p\} \int_{-\infty}^{\infty} |t| K_h(t) {\rm{d}}t \\
		=& h \max\{p,1-p\} \int_{-\infty}^{\infty} |t| K(t) {\rm{d}}t,
	\end{align}
	where the first inequality applies that $K_h(t)\ge 0$ and $\rho_p(x+t) - \rho_p(x) \le \big|\rho_p(x+t) - \rho_p(x) \big| $ for all $t$, the second inequality uses the property of piecewise linear function $\big|\rho_p(x+t) - \rho_p(x) \big| \le \max\{p,1-p\} |t|$, and the final equality uses $K_h(x)=\frac{1}{h} K \left(\frac{x}{h}\right)$ and variable substitution. So we have
	\begin{multline} \label{eq: conv_upper_bound}
		\tilde{f}_h^{\mathrm{conv}}(x)-f(x) = \sum_{i=1}^n \left[ l_p^h \big(s_i-x \big)-\rho_p \big(s_i-x \big)\right]
		\\ \le  n h \max\{p,1-p\} \int_{-\infty}^{\infty} |t| K(t) {\rm{d}}t, \forall x \in \mathbb{R}.
	\end{multline}
	Combining Eqs. \eqref{eq: conv_lower_bound} and \eqref{eq: conv_upper_bound}, we get $U_h=n h \max\{p,1-p\} \int_{-\infty}^{\infty} |t| K(t) {\rm{d}}t$.

{\color{black}
\section{Extension to multiple data points within each node} \label{extension:Ngn}

Considering the case where the $i$-th node has $n_i \ge 1$ scores $\{s_{i,j}\}_{j=1}^{n_i}$, our goal is to select the top-$k$ data points from $N=\sum_{i=1}^n n_i$, $i=1,\ldots,n$. The objective function of this problem becomes
	\begin{equation}
		\theta_k = \arg \min_{x\in \mathbb{R}} f(x) =\sum_{i=1}^n f_i(x) =\sum_{i=1}^n \sum_{j=1}^{n_i} \rho_p \big(s_{i,j}-x \big),
	\end{equation}
where $f_i(x)=\sum_{j=1}^{n_i} \rho_p \big(s_{i,j}-x \big)$ and $p\in (\frac{N-k}{N},\frac{N-k+1}{N})$. Actually, this problem with $n_i \ge 1$ is equivalent to the problem with $n_i = 1$, which only rearranges the data points and increases the number of data points from $n$ to $d$. Therefore, the analysis and algorithm are also similar with $n_i=1$, $i=1,\ldots,n$.  Using the same smoothing technique as described in the manuscript, we have
	\begin{equation}
		\theta_k = \arg \min_{x\in \mathbb{R}} \tilde{f}_h(x) =\sum_{i=1}^n \tilde{f}_i^h(x) =\sum_{i=1}^n \sum_{j=1}^{n_i} \rho_p^h \big(s_{i,j}-x \big),
	\end{equation}
where $\tilde{f}_i^h(x)= \sum_{j=1}^{n_i} \rho_p^h \big(s_{i,j}-x \big)$. The Algorithm 1 also works for this case. Besides, we note that for fixed $n$, the number of transmissions in each round doesn't change since $\nabla \tilde{f}_i^h(x)$ is computed locally. The analysis is without loss of generality, but the performance of our algorithm will improve, as the agents engage in communication by sharing only quantile estimates rather than raw data.}

\vspace{-0.5in}


\begin{IEEEbiography}[{\includegraphics[width=1in,height=1.25in,clip,keepaspectratio]{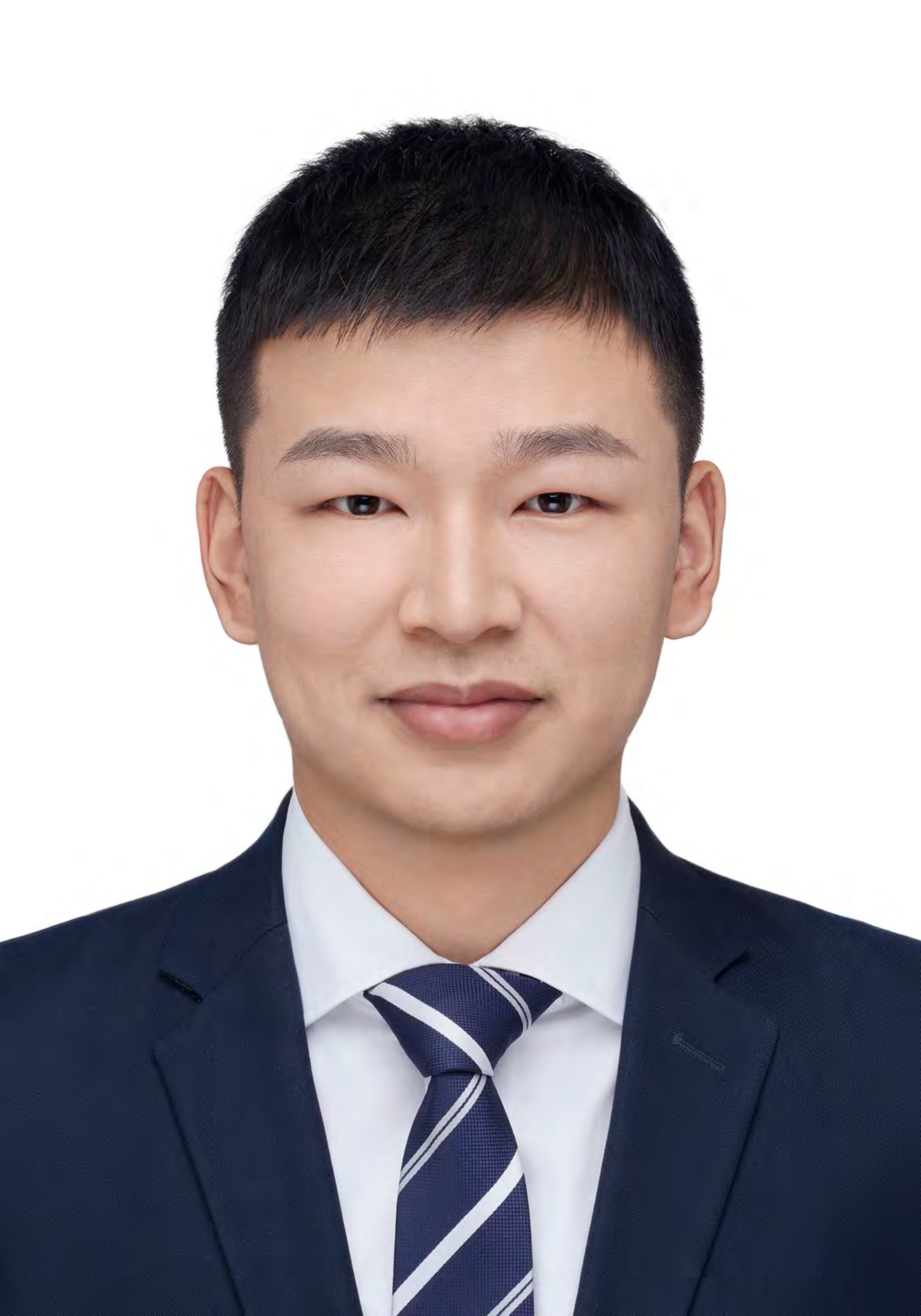}}]{Xu Zhang} is an assistant professor at the School of Artificial Intelligence, Xidian University. Previously, he was a Postdoctoral Researcher at the Academy of Mathematics and Systems Science, Chinese Academy of Sciences, from 2021 to 2023. He received his B.S. degree and Ph.D. degree in Electronics Engineering from the School of Information and Electronics, Beijing Institute of Technology, Beijing, China, in 2015 and 2021. He was a visiting student in the Ming Hsieh Department of Electrical Engineering at the University of Southern California from 2018 to 2019. His research interests include federated learning, distributed optimization, and inverse problems.
\end{IEEEbiography}	


\begin{IEEEbiography}[{\includegraphics[width=1in,height=1.25in,clip,keepaspectratio]{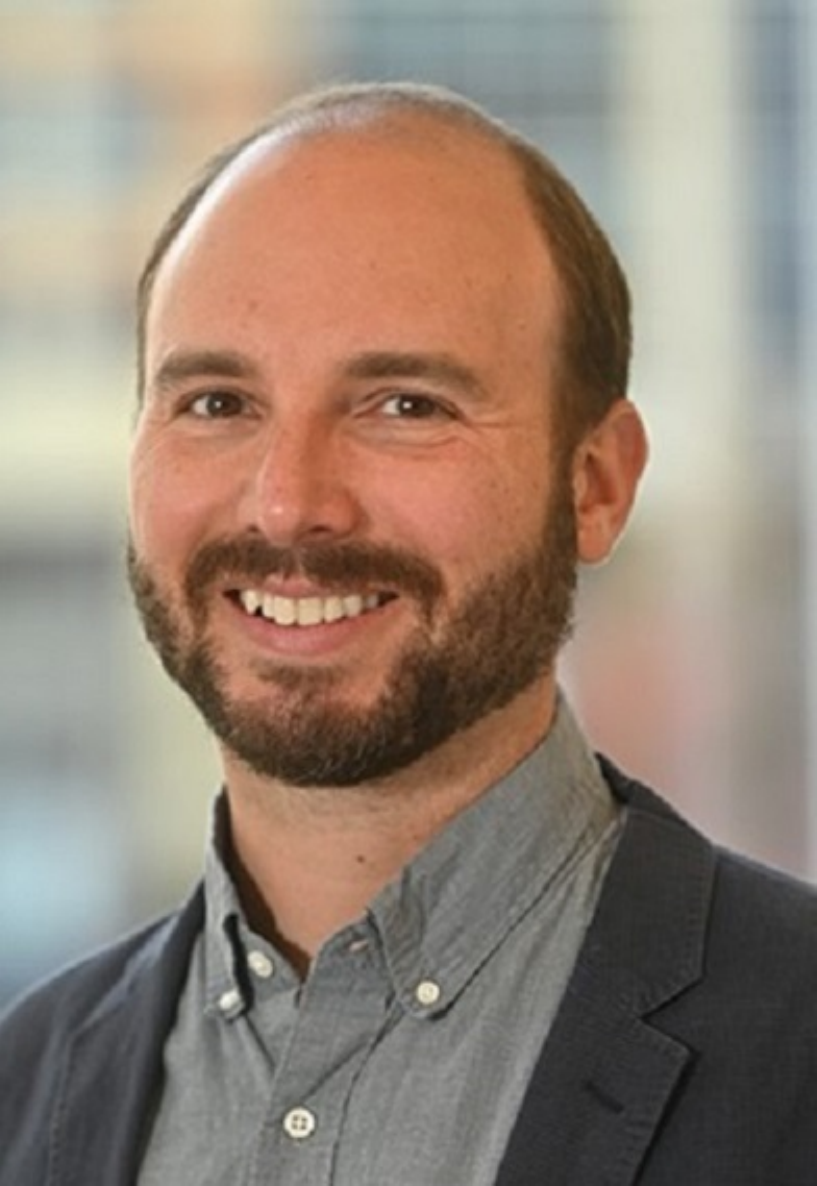}}]{Marcos M. Vasconcelos} is an Assistant Professor with the Department of Electrical Engineering at the FAMU-FSU College of Engineering, Florida State University. He received his Ph.D. from the University of Maryland, College Park, in 2016. He was a Research Assistant Professor at the Commonwealth Cyber Initiative and the Bradley Department of Electrical and Computer Engineering at Virginia Tech from 2021 to 2022. From 2016 to 2020, he was a Postdoctoral Research Associate in the Ming Hsieh Department of Electrical Engineering at the University of Southern California. His research interests include networked control and estimation, multi-scale robotic networks, game theory, distributed optimization, distributed machine learning, and systems biology.
\end{IEEEbiography}

\end{document}